\documentclass[sigplan,10pt]{acmart}

\AtBeginDocument{%
  }

\copyrightyear{2027}
\acmYear{2027}
\setcopyright{cc}
\setcctype{by}
\acmConference[EuroSys '27]{22nd European Conference on Computer Systems}{April 19--23, 2027}{Rabat, Morocco}
\acmBooktitle{22nd European Conference on Computer Systems (EuroSys '27), April 19--23, 2027, Rabat, Morocco}
\acmDOI{10.1145/3842654.3848541}
\acmISBN{979-8-4007-2971-3/2027/04}

\usepackage{array}
\usepackage{hyperref}
\usepackage{graphicx}
\usepackage{tabularray}
\usepackage{subcaption}
\usepackage{adjustbox}
\usepackage{algorithm}
\usepackage{multirow}
\usepackage{pgf-pie}
\usepackage{amsfonts}
\usepackage{algpseudocode}
\usepackage{listings}
\usepackage[svgnames, dvipsnames]{xcolor}
\usepackage{tikz}
\usetikzlibrary{calc}
\usetikzlibrary{shapes}
\usepackage{xspace}
\usepackage{amsmath}
\usepackage{amsthm}
\usepackage{soul}
\usepackage{float}
\floatstyle{plaintop}
\restylefloat{table}
\usepackage[normalem]{ulem}
\usepackage{enumitem}
\usepackage{booktabs}
\usepackage{tabularx}
\usepackage{etoolbox}
\usepackage{pifont}
\usepackage{natbib}
\usepackage[justification=justified]{caption}

\usepackage{hyperref}
\usepackage{xparse}

\newcommand{\fixme}[1]{#1}

\newif\ifshowrevisions
\showrevisionsfalse

\newcounter{revcomment}
\newcounter{revrevision}[revcomment]

\newcommand{\revtag}[1]{\textbf{[#1]}}

\newlength{\revsectionindent}
\newlength{\revitemleftindent}
\newlength{\revitembulletwidth}
\ExplSyntaxOn

\seq_new:N \l__revitem_tags_seq

\NewDocumentCommand{\revlabel}{m}
{
  \ifshowrevisions
    \hypertarget{revtext:#1}{}%
    \hyperlink{revitem:#1}{\revtag{\ref*{revnum:#1}}}%
  \else
    \hypertarget{revtext:#1}{}%
  \fi
}

\cs_new_protected:Npn \__rev_register_one_label:n #1
{
  \refstepcounter{revrevision}
  \label{revnum:#1}
}

\cs_new_protected:Npn \__rev_make_one_item_anchor:n #1
{
  \hypertarget{revitem:#1}{}
}

\cs_new:Npn \__rev_item_to_text_tag:n #1
{
  \hyperlink{revtext:#1}{\textbf{\ref*{revnum:#1}}}
}

\cs_new_protected:Npn \__rev_print_item_tag_list:n #1
{
  \seq_clear:N \l__revitem_tags_seq

  \clist_map_inline:nn {#1}
  {
    \seq_put_right:Nn \l__revitem_tags_seq
    {
      \__rev_item_to_text_tag:n {##1}
    }
  }

  \textbf{[}\seq_use:Nn \l__revitem_tags_seq { \textbf{,}~ }\textbf{]}
}

\NewDocumentCommand{\revitem}{mm}
{
  \clist_map_inline:nn {#1}
  {
    \__rev_register_one_label:n {##1}
  }

  \ifshowrevisions
    \par\noindent
    \clist_map_inline:nn {#1}
    {
      \__rev_make_one_item_anchor:n {##1}
    }%
    \hangindent=\dimexpr\revitemleftindent+\revitembulletwidth\relax
    \hangafter=1
    \hspace*{\revitemleftindent}%
    \makebox[\revitembulletwidth][l]{$\bullet$}%
    \ignorespaces#2~\__rev_print_item_tag_list:n {#1}%
    \par
  \fi
}

\ExplSyntaxOff

\newif\ifshowmanualdiff

\newif\ifrevise
\newif\ifrevisenew
\newif\iffinal
\newif\iffinalnew

\revisetrue
\revisenewtrue
\finaltrue
\finalnewtrue

\revisefalse
\revisenewfalse
\finalfalse
\finalnewfalse

\showrevisionstrue
\showmanualdifftrue

\showrevisionsfalse
\showmanualdifffalse

\ifshowmanualdiff
\newcommand{\mao}[1]{\textcolor{blue}{#1}}
\else
\newcommand{\mao}[1]{#1}
\fi

\newcommand{\ext}{\texttt{sched\_ext}~}
\newcommand{\sys}{\mbox{\textsc{Venus}}\xspace}
\newcommand*\WC[1]{\tikz[baseline=(char.base)]{
            \node[shape=circle,draw,inner sep=.5pt] (char) {#1};}}

\definecolor{ao}{rgb}{0.0, 0.5, 0.0}

\newcommand{\PreserveBackslash}[1]{\let\temp=\\#1\let\\=\temp}
\newcolumntype{C}[1]{>{\PreserveBackslash\centering}p{#1}}
\newcolumntype{R}[1]{>{\PreserveBackslash\raggedleft}p{#1}}
\newcolumntype{L}[1]{>{\PreserveBackslash\raggedright}p{#1}}

\algblock{Input}{EndInput}
\algnotext{EndInput}
\algblock{Output}{EndOutput}
\algnotext{EndOutput}

\definecolor{superlightgray}{rgb}{0.93, 0.93, 0.93}
\newcommand{\takeaway}[2]{%
\noindent %
\colorbox{superlightgray}{\begin{minipage}{.98\columnwidth}
\textbf{#1:} #2
\end{minipage}}
}

\newcommand*\sqbox[1]{\tikz[baseline=(char.base)]{
            \node[shape=rectangle,draw,inner sep=1.5pt] (char) {#1};}}

\begin{document}
\pagenumbering{arabic}


\title{No Provenance Left Behind: Revisiting Reference Monitor Guarantees with Learned Kernel Scheduling}


\author{Jinsong Mao}
\affiliation{%
  \institution{University of Massachusetts Amherst}
  \country{USA}
  }

\author{Benjamin E. Ujcich}
\affiliation{%
  \institution{Georgetown University}
  \country{USA}
  }

\author{Shiqing Ma}
\affiliation{%
  \institution{University of Massachusetts Amherst}
  \country{USA}
  }

\begin{abstract}
Provenance plays a critical role in maintaining traceability of a system's actions for root cause analysis of security threats and their impact.
Provenance collection is often implemented within system reference monitors to ensure that an audit trail exists of all events, that events are completely captured, and that logging of such events cannot be bypassed.
However, recent research has shown that existing state-of-the-art provenance collection systems can fail to ensure the security guarantees of a true reference monitor due to the ``super producer threat'' in which provenance generation can overload a system to force the system to drop security-relevant events and allow an attacker to hide their actions.
One approach towards solving this threat is to enforce resource isolation, but that does not fully solve the problems resulting from hardware dependencies and performance limitations.

In this paper, we show that an operating system's kernel scheduler can \mao{\revlabel{abstract-empirical}empirically mitigate} this threat, and we introduce \sys, a learned scheduler for Linux specifically designed for provenance.
Unlike conventional schedulers that ignore provenance completeness requirements, \sys leverages reinforcement learning to learn provenance task behavior and to dynamically optimize resource allocation.
We evaluate \sys's efficacy and show that \sys significantly improves both the completeness and efficiency of provenance collection systems compared to traditional scheduling, while maintaining reasonable overheads and even improving overall runtime in certain cases compared to the default Linux scheduler.
\end{abstract}



\maketitle

\section{Introduction}\label{sec:intro}

The global average cost of a data breach in 2024 marked a 10\% increase from 2023, reaching an all-time high\cite{ibmCostData}, underscoring the critical need for robust defense and forensics.
Provenance systems address this need by systematically capturing and tracing the origins and pathways of data flows, offering invaluable insights into the evolution of data over time.
This functionality is essential for applications such as forensic analysis and intrusion detection.
By recording detailed interactions within computing environments, provenance systems promote transparency and accountability, even in highly complex scenarios.
The reliability and effectiveness of these systems, however, hinge on the \emph{completeness} of captured provenance events.
In other words, provenance systems should follow the reference monitor design~\cite{anderson1972computer}, ensuring that \emph{all} system events are captured and logged.

\smallskip
\noindent
\textbf{Revisiting  reference monitor guarantees~~}
We revisit the reference monitor concept in the context of provenance systems and find that existing solutions do not adequately guarantee the completeness of provenance events.
In particular, the ``super producer threat''~\cite{nodrop} in which provenance event generation overloads the system and causes losses of security-relevant events, poses a significant challenge to the reference monitor guarantees.
This attack is surprisingly effective and easy to implement, as it does not require any special privileges or knowledge of the provenance system.

Recent solutions, such as NoDrop~\cite{nodrop}, propose 
resource isolation so that the host process consumes its own generated events to prevent provenance event loss.
Such losses are curtailed by injecting consumer code into the host process and enforcing its execution to clean the filled event buffer.
Unfortunately, such resource isolation is challenging to deploy: it requires specialized hardware support (i.e., hardware-based Intel Memory Protection Keys), it is tightly coupled to specific legacy kernels, 
and it introduces issues when executing atomic operations that may lead to errors (e.g., \texttt{Scheduling while atomic}) and potential system crashes.
Additionally, the enforcement of injected consumer code may introduce performance overheads and fairness issues. 

\smallskip
\noindent\textbf{Our insight~~}
The key to mitigating super producer threats is not detection, but proper resource allocation. The threat exists when
the provenance system does not consume the generated events in time, leading to buffer overflows and event loss.
Our insight is that a kernel scheduler can \fixme{mitigate} this problem by ensuring the provenance system is allocated sufficient resources to consume generated events in a timely manner.
However, tackling this challenge is not straightforward.
Traditional schedulers, such as Linux's default schedulers (e.g., the CFS~\cite{cfs} and the EEVDF~\cite{eevdf} scheduler), are designed to provide a generic and balanced scheduling experience optimized around performance metrics like throughput, fairness, and latency among generic workloads, but they lack an understanding of provenance systems'  unique characteristics and requirements.
This deficiency can result in lost provenance, which compromises provenance's reliability and security and the overall system's reference monitor guarantees.

\smallskip
\noindent\textbf{Overview~~}
The design of \sys is driven by three primary goals: completeness, efficiency, and fairness.
\mao{\revlabel{intro-empirical}Completeness refers to whether provenance events are captured without observed loss}, even under extreme workloads, safeguarding the integrity and reliability of provenance systems.
Efficiency aims to maintain or improve system performance by optimizing resource allocation while minimizing computational overhead.
Fairness ensures balanced resource distribution, preventing starvation and maintaining proportional CPU utilization across diverse workloads.

To achieve these goals, \sys introduces a kernel-space learned scheduler tailored to the unique demands of provenance systems.
The method combines a queue-based scheduling framework with reinforcement learning, offering dynamic adaptation and resource optimization.

At a high level, \sys organizes tasks into multiple queues, each equipped with predefined waiting times and resource budgets.
Provenance-related tasks are allocated to non-primary queues, which enforce strict resource constraints, while general tasks are managed by a primary queue.
By dynamically selecting tasks from the most ``hungry'' eligible queue, \sys ensures fair and proportional resource allocation, preventing starvation and maintaining system stability.

\sys is powered by a lightweight Deep Q-Network (DQN) that predicts the optimal scheduling decisions based on task behavior and system context.
This reinforcement learning model analyzes features, such as event generation rates, buffer availability, and task runtime patterns, to dynamically adjust queue placement.
Dual rewards underpin the learning process: a provenance reward penalizes event loss to ensure provenance completeness, while a utilization reward minimizes CPU idleness to enhance overall efficiency.
These rewards enable \sys to learn and adapt to diverse workloads, addressing the challenges posed by unpredictable and bursty event streams.
To further improve efficiency, \sys incorporates a delta function to skip unnecessary scheduling decisions when tasks are stable, reducing computational overhead.
It also employs an exponential waiting time strategy for queue consumption, ensuring predictable scheduling and precise resource control.

To ensure the security and stability of the extended kernel functionality,
\sys is implemented within the Linux kernel using the eBPF subsystem and the \texttt{sched\_ext} framework.
Both provide strong guarantees through rigorous verification and controlled kernel extensions.
This kernel-space implementation also avoids unnecessary data transfer and context switching, collecting provenance and task features directly from kernel data structures.
By integrating core components into the kernel, \sys ensures compatibility with modern Linux versions and minimizes integration complexity.

\noindent\textbf{Contributions~~}
We evaluate \sys on eight benchmarks and two provenance systems, comparing it with three other schedulers.
\mao{\revlabel{contribution-empirical}
Although \sys does not formally guarantee event completeness, our empirical}
results show that \sys prevents observed provenance event loss in settings where other schedulers lose 39\% to 98\% of events.
Meanwhile, \sys improves average macrobenchmark runtime by 3.42\%, 0.90\%, and 0.25\% compared with EEVDF\cite{eevdf}, LAVD\cite{lavd}, and Rusty\cite{rusty}, respectively.
Our contributions are summarized as:
\begin{itemize}[noitemsep,topsep=0pt,leftmargin=12pt]
\setlength\itemsep{0em}
    \item[1.] We revisit the long-standing reference monitor concept in the context of provenance systems and discover that the super producer threat, which leads to the loss of provenance events, poses a significant challenge to the reference monitor guarantees even in state-of-the-art systems.
    \item[2.] We propose and implement a novel reinforcement-learning-based scheduler, \sys, that achieves
    \mao{\revlabel{contribution-empirical2}the desired empirical security and performance properties}.
    To the best of our knowledge, \sys is the first machine-learning-based kernel scheduler. Our source code is available at \url{https://github.com/4ar0nma0/Venus} .
    \item[3.] We evaluate \sys with macro and micro benchmarks, illustrating its superior effectiveness and performance compared to existing schedulers.
\end{itemize}

\section{Background \& Challenges}

\noindent\textbf{Provenance~~}
Provenance systems serve as the foundation for understanding the lifecycle of data within computational systems. These systems capture, record, and analyze detailed information about the origins, transformations, and consumption of data.
Provenance data is critical for various applications, ranging from system auditing\cite{audit1, kairos, eaudit, bates2017transparent, inam2023sok}, forensic analysis\cite{forensics1, forensics2, forensics3, forensics4, forensics5}, data security\cite{ds1,ds2,ds3,atlas,airtag} and reproducibility\cite{rain, recprov,rr1,rr2,rr3}. For example,
by recording system events,
provenance systems enable the reconstruction of attack vectors and identification of malicious activities\cite{atlas,kairos,airtag, causal1, causal2,causal3,causal4,loggc,nodemerge}.

Provenance systems are typically structured around  \emph{producers} and \emph{consumers}.
Producers use hooks integrated across various layers of the software stack that enable the capture of a wide range of activities.
For instance, hooks embedded within the operating system kernel can capture system calls such as
\texttt{open} for file access or
\texttt{execve} for process creation and execution.
Consumers, in contrast, receive, store, and analyze the produced provenance data. Consumers can range from simple storage mechanisms
to sophisticated online analytics engines.
However, producers and consumers share a common challenge: efficient and provenance-aware scheduling mechanisms to ensure no events are lost.

\vspace{4pt}
\noindent\textbf{Reference monitors~~}
Reference monitors~\cite{anderson1972computer} are a fundamental security mechanism in operating system design that enforce system-critical security policies (e.g., access control) by acting as an interposer on every access attempt between subjects (e.g., users or processes) and objects (e.g., files, devices).
They provide the properties of being non-bypassable, evaluable, always invoked, and tamper-proof.

Most existing system-level provenance solutions\cite{sysdig, eaudit, nodrop, tetragon, tracee, lpm} embed provenance collectors within the overall system's kernel-space reference monitor like Linux Provenance Modules\cite{lpm} or eBPF subsystem to ensure that the access decisions, along with their context (e.g., who accessed what and why), are recorded securely and cannot be bypassed, with the reference monitor serving as part of the trusted computing base (TCB). This enhances the overall system's security and provides the basis for auditing.

\subsection{Case Study: Super Producer Threat}\label{sec:moti}

We consider a representative vulnerability of a provenance system within a reference monitor when confronted with the ``super producer threat''~\cite{nodrop} in which an adversary attempts to defeat the reference monitor security guarantees.
We devise a representative example using Sysdig \cite{sysdig}, a widely used open-source provenance system, and the default Linux EEVDF scheduler.
Sysdig should continuously collect system events and save the provenance data to the local disk.

\noindent\textbf{Attack~~}
We assume that an adversary aims to exploit scheduling weaknesses to mask malicious events.
The adversary  deploys a ``super producer,'' which generates a significant volume of system calls.
While the super producer is running, the adversary compresses a secret folder, \texttt{secret}, containing four files (\texttt{file1} to \texttt{file4}), into an archive, \texttt{secret.zip}. The compressed archive is then uploaded to a remote server at \texttt{1.2.3.4} using \texttt{curl}. Subsequently, the adversary downloads a script, \texttt{suspicious.sh}, from the same host and overwrites a previously benign script, \texttt{maintenance.sh}, with this malicious payload.
The numerous events triggered by the super producer overwhelm the consumer buffer, leading to missed events associated with these malicious operations.

\begin{figure}
    \centering
    \subfloat[Sysdig suffers from inadequate scheduling and resource allocation. 
    The line indicates the percentage of provenance events loss.
    ]{
    \includegraphics[width=.95\linewidth]{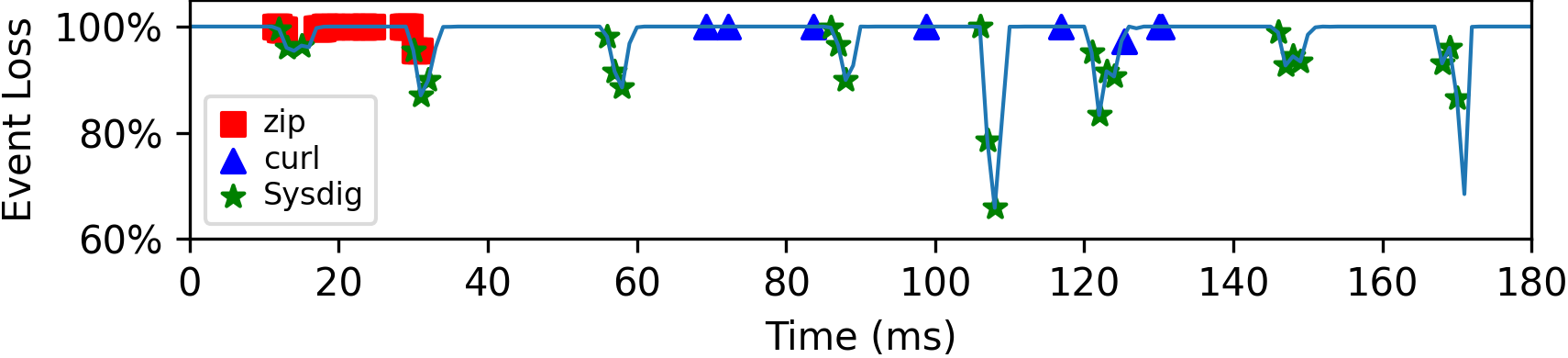}
    \label{fig:moti-example}
    }
    
    \subfloat[
    Causal graphs generated during attack investigations.
    With a super producer, the adversary hides evidence of malicious events since such events were not captured.
    ]{
    \includegraphics[width=.95\linewidth]{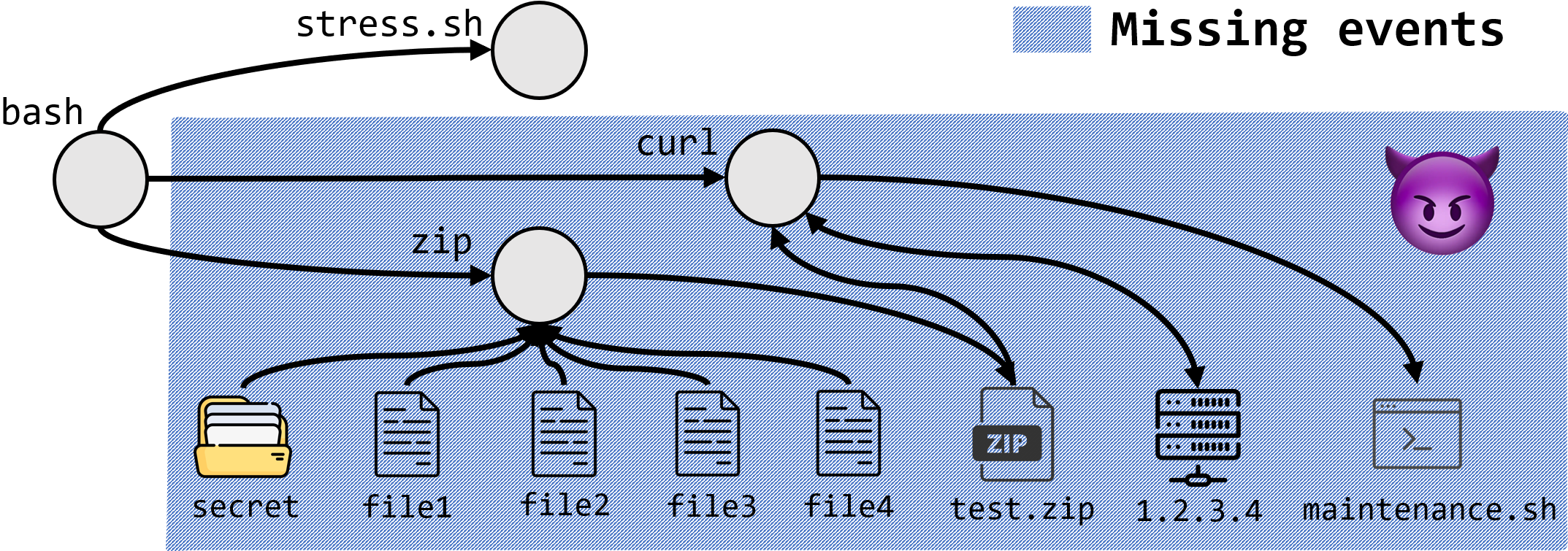}
    \label{fig:moti-causal}
    }
    \caption{Scheduling statistics and causal graphs of the Sysdig and EEVDF ``super producer''~\cite{nodrop} case study example.}
    \label{fig:moti}
\end{figure}

\noindent\textbf{Results~~}
\autoref{fig:moti-example} illustrates the event loss statistics and the scheduling order of all programs during the malicious operations. The blue line represents the average number of events lost over 1 ms intervals. Stars, squares and triangles indicate when a corresponding program is scheduled.
The results reveal that the consumer was scheduled too infrequently and the system experienced a significant loss of provenance events during the super producer due to buffer overflow. All the malicious workloads were executed in the context of near 100\% event loss, indicating the provenance system is not recording these malicious operations.

\autoref{fig:moti-causal} shows the causal graphs generated during attack investigation. Normally the provenance system collects all events, and the causal graph shows that the adversary who invoked \texttt{stress.sh} also visited \texttt{secret} and its files. The upload and consequent injection to \texttt{maintenance.sh} are also shown. However, due to an incomplete provenance log, the records of these malicious operations are lost and the adversary successfully hides the malicious events.

\subsection{Defense Challenges}\label{sec:challenges}

We identify what makes the super producer threat challenging to defend against with existing approaches.

\noindent\textbf{Low-cost barrier to entry~~}
The attack requires few resources for an adversary to perform effectively. In our experiment, the stress imposed on the provenance system was minimal, achieved by scanning the \texttt{/usr} directory with just five concurrent threads in addition to data collection. This type of operation can be invoked by any user at any time, regardless of the machine type or scenario, making it easy to execute.
Furthermore, the attack is also stealthy: it operates independently and is not directly linked to any subsequent malicious actions or external software. As a result, such activity is more likely to be misclassified as benign.

\noindent\textbf{Limitations of existing defenses~~}
Based on known prior work, we find that nearly all state-of-the-art provenance systems suffer from this attack, and state-of-the-art solutions fail to fully address the problem.

We measured three key aspects of a provenance system that are important to the attack: {\textit{completeness}} (whether the system ensures no provenance event loss or is capable of performing mitigation);
{\textit{budget-aware}} (whether the system is aware of resource allocation and makes efficient use of it); and
{\textit{compatibility}} (whether the solution can perform regardless of software, hardware, and provenance-system requirements).
We find the following observations and trends:

\begin{itemize}[noitemsep,topsep=0pt,leftmargin=*]
\setlength\itemsep{0em}
\item Nearly all modern provenance systems\cite{eaudit,sysdig,tetragon,tracee,auditd,camflow} suffer from event loss, which
undermines the reference monitor security guarantees of a provenance system.
\item
\mao{\revlabel{eaudit}
eAudit\cite{eaudit} and HardLog\cite{hardlog} are partially budget-aware. However, their core designs rely on encoding, multiple buffers, and event prioritization. Without sufficient scheduling resources, event production can exceed consumption and ultimately lead to buffer overflow.
}
\item Some systems require specific auditing devices~\cite{hardlog,omnilog}, software that lives outside of the standard kernel source tree~\cite{eaudit}, or hardware-based defenses~\cite{nodrop}, which limits wider deployment.
\end{itemize}

\begin{figure}
    \centering
    \includegraphics[width=\columnwidth]{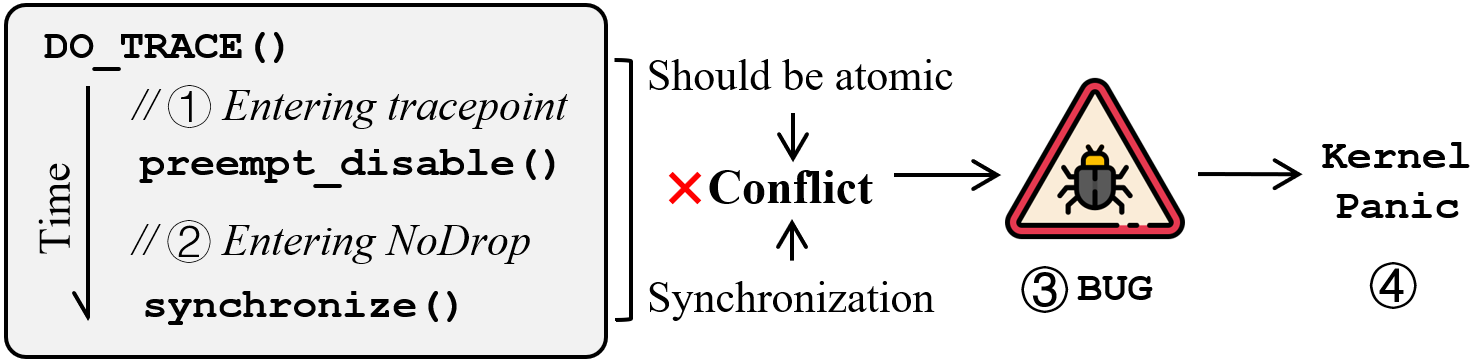}
    \caption{NoDrop needs to perform  synchronization in critical regions, causing kernel bugs, deadlocks and potential kernel panic~\cite{freeze1, panic1,panic2}.}
    \label{fig:nodrop}
\end{figure}

\mao{\revlabel{nodrop-omnilog-comparison}
The most relevant works are OmniLog\cite{omnilog} and NoDrop\cite{nodrop}. OmniLog provides high event coverage, but focuses on whether security-relevant events are generated for attack investigation rather than whether generated events can always be collected and stored under resource pressure.
NoDrop achieves event completeness by introducing per-thread resource isolation for event processing and storage.
However, NoDrop faces practical compatibility constraints. As shown in~\autoref{fig:nodrop}, on mainstream kernels its design requires synchronization inside a tracepoint, which executes in an atomic section. Such synchronization can trigger scheduler timeouts and forced rescheduling, leading to potential deadlocks or kernel panics~\cite{freeze1,panic1,panic2}. In addition, its resource isolation mechanism relies on Intel Memory Protection Keys, limiting deployment to supported hardware platforms. Moreover, NoDrop is a provenance-system-level design rather than a general defense, and therefore does not directly address event completeness for other provenance systems.
}

Ideally, we want a system that is free from hardware constraints, is portable, is performant and works seamlessly with the kernel.
We consider the \emph{kernel scheduler} as a way to overcome those challenges, which offers fundamental and universally available mechanisms for resource allocation.

\section{\sys Goals}\label{sec:goal}

\label{sec:requirements}

\smallskip
\noindent\textbf{Threat model~~} We assume that the scheduler and the provenance system are trusted and that the scheduler is capable of accessing statistics from the provenance system. \mao{\revlabel{threat-model}As in prior work\cite{mithridates,hassan2020tactical}, we assume a trusted preparation window; after this window,} any user can invoke any userspace workload that does not harm kernel space, including a potential super producer invoked by an adversary. The security of the transmission and storage of provenance logs is well-studied\cite{hardlog, prov-storage1, prov-storage2, prov-storage3, prov-storage4, seal, loggc, nodemerge, sealfsv2, sealfs} and beyond the scope of this paper.

\subsection{Design Goals and Desired Properties}
Based on the threat model, we consider the following design goals and properties of what a scheduler should provide.
First, we define a set of design goals that \sys should uphold to maintain the completeness and usability of provenance systems, particularly under adversarial or high-load conditions. Second, we outline a set of desired properties that, while not strictly required, enhance the performance and stability of \sys in diverse environments.

\noindent\textbf{Design Goals~~} Since the attack originates from corrupting the availability of provenance events, the completeness of collected provenance is the scheduler's top priority. At the same time, \sys must prevent starvation. We propose:

\begin{itemize}[noitemsep,topsep=0pt,leftmargin=19pt]
    \item[G1.]\textbf{\textit{\mao{\revlabel{goal-empirical}Empirical} overload prevention:}}
    The scheduler must have the capability to restrict the tasks' resource allocations (\autoref{sec:backbone}). This capability must be applied effectively to postpone or throttle jobs that risk exceeding the system's safe capacity. (\autoref{sec:rl})

    \item[G2.] \textbf{\textit{Quality of service:}}
    All tasks scheduled by \sys must be guaranteed to complete execution within a predictable time, ensuring consistent forward progress. (\autoref{sec:rl})
\end{itemize}

\noindent\textbf{Desirable Properties~~} Additionally, the scheduler should minimize the performance impact on the operating system:
\begin{itemize}[noitemsep,topsep=0pt,leftmargin=19pt]
    \setlength\itemsep{0em}
    \item[G3.] \textbf{\textit{Fairness:}} Each runnable process should receive a proportional share of CPU time over the long run, preventing starvation and ensuring balanced resource allocation. (\autoref{sec:backbone})

    \item[G4.] \textbf{\textit{Throughput:}} The system should maximize the total amount of work completed in a given timeframe, increasing overall efficiency. (\autoref{sec:rl})

    \item[G5.] \textbf{\textit{Overhead:}} The computational costs associated with scheduling should be minimized, including decision making and regular maintenance. The scheduler should also minimize other resource consumption. (\autoref{sec:overhead})

\end{itemize}

For practical concerns, we also want \sys to be forward compatible and adaptable to different provenance systems.
By achieving these goals, the scheduler can help ensure that the provenance system meets the reference monitor security, particularly the properties of being non-bypassable, evaluable, and always invoked.

\subsection{Scheduling Solutions and Limitations}\label{sec:limitations}

\mao{\revlabel{why-RL}
To provide these properties, \sys requires both new scheduling controls and a policy that can adapt those controls to workload and provenance-system behavior.
}

\mao{
\smallskip
\noindent\textbf{Existing scheduler}
designs do not provide overload prevention capability. Nice values, scheduling weights, and time-slice adjustments can promote the provenance consumer, but they do not directly restrict producers that exceed the provenance system's safe capacity. As shown in~\autoref{fig:moti-example}, fairness-oriented mechanisms such as CFS and EEVDF may still delay the consumer during bursty workloads and fail to consume the provenance buffer in time.
}

\mao{
\smallskip
\noindent\textbf{Heuristics}
can be effective, but they often reduce throughput or require substantial operational tuning.
First, observed buffer pressure is delayed feedback: simple buffer-threshold-based policies may react too late when the threshold is high, or unnecessarily throttle benign workloads when the threshold is low.
Fine-grained policies are also operationally expensive. For example, a cgroup-based policy could restrict producer CPU usage, but choosing the right budget requires detailed knowledge of event-generation rates, consumer throughput, and buffer capacity. Maintaining such policies therefore requires repeated profiling, tuning, and revalidation.
These limitations motivate a policy that can learn producer-consumer behavior from richer runtime signals rather than relying on fixed rules.
}

\mao{
\smallskip
\noindent\textbf{\sys.}
To address these limitations, we propose \sys, which provides the missing overload-prevention capability and uses RL for adaptation. Prior works have shown that RL can outperform heuristics for real-world resource-management problems~\cite{deeprm,decima, rl-nonlinear,nn-approx}.
}

\mao{
\revlabel{RL-limitation}
RL does not provide a formal proof of event completeness, which is a limitation of \sys. We mitigate this limitation by using RL only to tune scheduling decisions within a bounded queue-based framework, and by empirically validating overload prevention across workloads and provenance systems in~\autoref{sec:eval}. Moreover, adversarial manipulation is difficult under our threat model: poisoning would require control over the trusted training environment, while evasion would require visibility into system-wide provenance state and the ability to predict kernel-space scheduling decisions. Therefore, \sys uses RL as a practical adaptation mechanism without expanding the attacker capabilities considered by our threat model.
}

We summarize the properties provided by \sys in \autoref{tab:scheduler}. Compared to MLFQ and MLQ, \sys provides a stronger fairness design aligned with CFS and EEVDF while avoiding starvation. At the same time, \sys provides overload prevention capability that CFS, EEVDF, and MLFQ lack. Together, the queue-based control mechanism and data-driven tuning make \sys a practical and performant scheduler for provenance systems.

\section{\sys Design}
\label{sec:design}

\autoref{fig:overview} shows the overall system design of \sys.
\sys uses a two-layer architecture. The first layer establishes the \emph{backbone} of \sys (\autoref{sec:scheduler}) with a complete scheduling framework  and foundational capabilities of overload prevention and fairness. 
The second layer tunes this backbone with specific \emph{optimization} goals (\autoref{sec:adaptation}), such as attack mitigation and throughput maximization, to enable \sys to adapt to the underlying machine and provenance system. Simultaneously, \sys implements measures to minimize scheduling overhead, ensuring that the enhancements do not introduce significant performance penalties. 

\begin{table}
\centering
\resizebox{\columnwidth}{!}{
\begin{tabular}{L{1.4cm}L{2.15cm}L{2cm}L{2.2cm}}
\toprule%
\raisebox{0.5\baselineskip}{\bfseries{Scheduler}} & 
\multicolumn{1}{c}{\shortstack{\bfseries{Fairness} \\ (CPU Share)}}  & 
\multicolumn{1}{c}{\shortstack{\bfseries{Starvation} \\ \bfseries{Guarantee}}} & 
\multicolumn{1}{c}{\shortstack{\bfseries{Overload} \\ \bfseries{Prevention}}} \\
\cmidrule[0.6pt](r{0.125em}){1-4}%

CFS\cite{cfs}     & Proportional to priority  & \textbf{\textcolor{ao}{\checkmark}~}Task runs eventually  & \textcolor{BrickRed}{\ding{56}~}No capping mechanism \\ [-2ex]\\
EEVDF\cite{eevdf}   & Proportional to priority  & \textbf{\textcolor{ao}{\checkmark}~}Task runs eventually  & \textcolor{BrickRed}{\ding{56}~}No capping mechanism \\ [-2ex]\\
MLFQ\cite{mlfq}    & Related to CPU and I/O usage   & \textbf{\textcolor{ao}{\checkmark}~}Task runs per fixed time  & \textcolor{BrickRed}{\ding{56}~}Uncapped by priority boost \\ [-2ex]\\ 
MLQ\cite{mlq}     & Related to priority  & \textcolor{BrickRed}{\ding{56}~}\,Starves due to domination  & \textbf{\textcolor{ao}{\checkmark}~}Capped by domination \\ [-2ex]\\ 
\sys    & Proportional to priority  & \textbf{\textcolor{ao}{\checkmark}~}Task runs per fixed time  & \textbf{\textcolor{ao}{\checkmark}~}Capped by waiting-time\\ [-2ex]\\

\bottomrule
\end{tabular}
}
\caption{
Properties of \sys compared to existing schedulers.
CFS / EEVDF are Linux default schedulers, whereas MLQ and MLFQ share \sys’s multi-queue architecture.
}
\label{tab:scheduler}
\end{table}

\noindent\textbf{Security Implications}
\sys preserves kernel safety and availability properties while supporting the paper's reference-monitor goals.
This is ensured by two mechanisms: eBPF and \ext.
All \sys components are implemented using eBPF, which enforces strict verification at load time, preventing unsafe or malicious code from entering the kernel\cite{verifier}. In addition, \ext provides a built-in fallback to the default scheduler (e.g., EEVDF), ensuring that the system can safely recover in the event of misbehavior or starvation, preserving system integrity and availability.

\subsection{\sys Backbone}\label{sec:scheduler}

Similar to the classic MLQ and MLFQ scheduling, a task under \sys is placed into an appropriate queue according to its behavior. However, \sys has two key ideas: \WC{1} its unique queue design provides overload prevention capability and fairness properties; \WC{2} it collects task runtime statistics and uses a neural network to dynamically assign tasks to queues.
We first discuss how \sys gains such capabilities, then show how \sys adapts to different workloads and provenance systems.

\subsubsection{
\sys queue design}\label{sec:backbone}

\sys  assigns tasks into different first-in-first-out (FIFO) queues with a fixed time slice. When a task becomes runnable, it is first moved to a dispatch queue. When the CPU requests \revlabel{typo2} a task to run, a queue is consumed (i.e., the first task in the selected queue is moved to the local run queue and is run). As a result, every task in the same queue has an equal chance to run.

\sys employs a \emph{primary queue} and several \emph{non-primary queues}. 
The primary queue, operating on a FIFO basis, provides a default scheduling mechanism: when all tasks are directed to this queue, \sys behaves like a traditional FIFO scheduler.
The overload prevention capability comes from non-primary queues.
Non-primary queues are also FIFO but with designated waiting times, which are designed as natural resource budgets. A non-primary queue becomes eligible for consumption only when the elapsed time since the last task execution in that queue exceeds its waiting time, which enforces a strict resource allocation policy and ensures that tasks within one non-primary queue share the same pre-allocated budget.
When all non-primary queues are not eligible to be consumed, the primary queue is consumed.

\smallskip
\takeaway{Takeaways}{
\sys is able to achieve a balanced and controlled production-consumption process.
That mitigates the risk of exceeding the system's safe capacity.
}

\begin{figure}[t]
    \centering
    \includegraphics[width=\columnwidth]{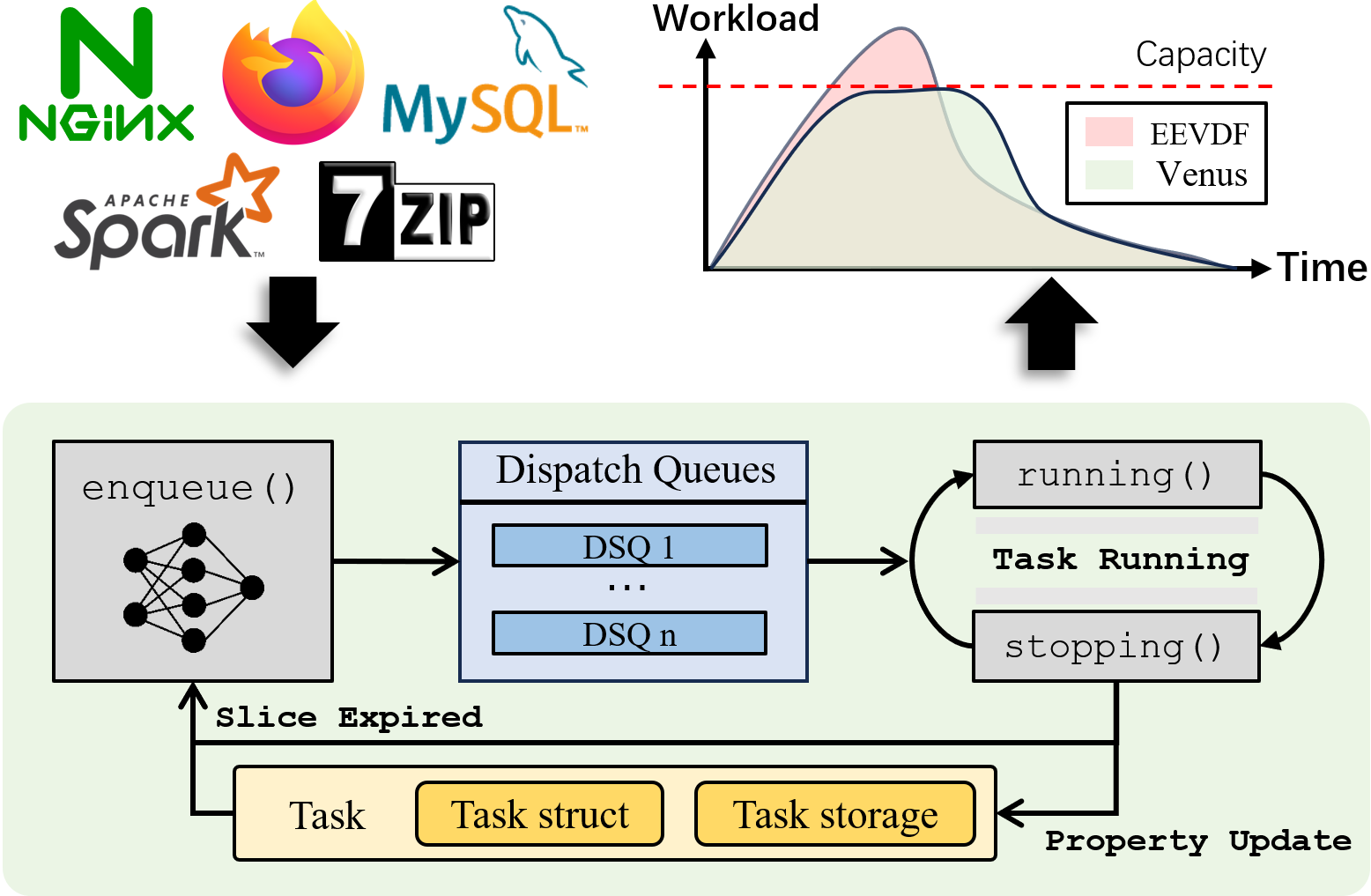}
    \caption{Overview of \sys, using task and provenance features to 
    \fixme{prevent event loss while preserving performance.}}
    \label{fig:overview}
\end{figure}

\smallskip
\noindent\textbf{Fairness~~}
To achieve fairness among all queues, when the CPU is idle and requests tasks to execute, \sys consumes the most ``hungry'' non-primary queues---those that have waited the longest since their last execution. If only the primary queue is eligible, the primary queue is consumed. This scheduling strategy ensures that queues operate in a fair and predictable manner.

Let $N$ be the number of queues, where queue 1 being the primary queue and queue 2 to N being the non-primary queues. Let $T=\{t_1,...,t_N\}\in \mathbb{N}^N$ be the time elapsed since the queues were picked and $\hat{T}=\{\hat{t}_2,...,\hat{t}_N\}\in \mathbb{N}^{N-1}$ be the waiting time of the queues. A queue $N$ is hungrier when $t_N$ is less and $\hat{t}_N$ is more. We define the hungry factor $h$ as:
\begin{equation}
\begin{split}
    h_i = \frac{t_i}{\hat{t}_i},~ i>1 ;\quad
    H &= \{h_1,...,h_N\}\in \mathbb{R}^N
\end{split}
\end{equation}
Specifically, $h_1=0$ because the primary queue is only consumed when no other queues are eligible, and thus, the least hungry.
Denote the number of tasks in the queues as $M=\{m_1,...,m_N\}\in \mathbb{N}^N$. \sys always chooses the hungriest eligible non-empty queue $k$ as shown in ~\autoref{eq:queues}:

\begin{equation}
    k=\text{argmax}_i(~h_i~|~h_i\geq 1,m_i>0)
\label{eq:queues}
\end{equation}

\smallskip
\noindent
\sys provides no-starvation guarantee and proportional resource allocation for non-primary queues. Let the waiting time be in increasing order $\hat{t}_2<...<\hat{t}_N$. In the worst case for the lowest priority queue $N$, \sys runs on a single-core system where low-priority queues have less chance to run than in a multi-core system, and the fixed time slice $s>\hat{t}_N$ so that high-priority queues will more likely compete with low-priority ones. We prove that the lowest priority queue, $N$, does not starve.
Given any state at the beginning, let the elapsed time be $t$. Queue 3 is consumed when $t/\hat{t}_3>s\cdot N/\hat{t}_2$, giving the guarantee of queue 3 as $t_3<s\cdot N\cdot \hat{t}_3/\hat{t}_2$. Similarly, we obtain the guarantee for queue $N$ as follows:
\begin{equation}
    t_N<\frac{s\cdot N \cdot \hat{t}_N}{\hat{t}_2},\quad s>\hat{t}_N>...>\hat{t}_2
\label{eq:starvation}
\end{equation}
This indicates the consumption time is bounded and the bound for each queue is proportional to its waiting time.

\sys provides finite execution guarantees. Following \autoref{eq:starvation}, the runtime of a task in queue $N$ is bounded as follows compared with round robin (RR) scheduler:
\begin{equation}
\begin{split}
    \text{runtime}_{\footnotesize \text{Venus}}  \leq \frac{\hat{t}_N\cdot N}{\hat{t}_2}  \cdot \text{runtime}_\text{\footnotesize RR}
\end{split}
\label{eq:finish_time}
\end{equation}
This indicates \sys ensures that even tasks in the lowest-priority queue finish in a reasonable time - proportional to the number of configured queues and the corresponding waiting time - compared to the classic RR scheduling. In real-world cases, the execution time is often significantly less than the bound, through per-machine adaptation.

\smallskip
\takeaway{Takeaways}{
By preventing starvation and achieving proportional CPU share, tasks complete in the long run with queue-based fairness.
}

\subsubsection{Understanding the adaptation space}
The backbone leaves an adaptation space consisting of the neural network and the waiting time configuration.
\smallskip

\noindent\textbf{Neural network and features~~}
An important question is which queue to put a specific task in. The behaviors of tasks can vary, and it is difficult for traditional methods to model and optimize over tasks, especially with different provenance configurations. In \sys, we address the issue with a neural network (NN). An NN can be extensibly deployed to predict the appropriate queue for the task, as long as the network is well trained and adapts to the current configuration.

We configure the neural network to make queue predictions according to the context of the task, which describes its recent behavior and serves as input of the neural network. We set up the context with
    \emph{task features} that describe how the task behaves and interacts with other tasks and 
    \emph{provenance features} that describe how the task produces events and the status of the provenance system.
A detailed overview of the context is shown in ~\autoref{tab:feature}.\footnote{\texttt{wait\_freq} and \texttt{wake\_freq} are adopted from the LAVD\cite{lavd} scheduler; other features can be obtained from \texttt{task\_struct} or the provenance system. }
Together, these features provide comprehensive information about the task's nature and impact, and the provenance system's current state.

\smallskip
\label{sec:waiting-time}
\noindent\textbf{Waiting time~~}
Potential combinations of waiting time settings lead to an impractical search space. To address this, we narrow the searching space based on the stress tests.
We subject the provenance system to peak workloads by manually generating as many system calls as possible. In this process, two queues are configured, where a non-primary queue runs all stress workloads, and the primary queue runs all other tasks, including the provenance consumer. We then incrementally increase the waiting time until the system reaches a stationary state where no events are lost.

\begin{table}
\centering
\footnotesize
\resizebox{\linewidth}{!}{
\begin{tabular}{C{1cm}C{1.5cm}L{4cm}}
\toprule%
\bfseries{Category} & \bfseries{Metric} & \multicolumn{1}{c}{\bfseries{Description}} \\
\cmidrule[0.6pt](r{0.125em}){1-1}%
\cmidrule[0.6pt](lr{0.125em}){2-2}%
\cmidrule[0.6pt](lr{0.125em}){3-3}%

\multirow{7}{*}{Task}  
& \texttt{runtime}      & Total CPU time used by this task \\
& \texttt{nvcsw}        & Avg.\# voluntary context switches\\ 
& \texttt{wait\_freq}   & Freq. of task waits precedents\\ 
& \texttt{wake\_freq}   & Freq. of task wakes subsequences \\  
& \texttt{nr\_stop}     & \# task's stops in current slice \\
& \texttt{nr\_idle}     & \# idle slides before this task  \\
& \texttt{prio}         & The task's current queue  \smallskip  \\ 

\hline
\\[-1.5ex]

\multirow{4}{*}{Provenance}
& \texttt{nr\_event}    & Avg.\# events generated by the task \\
& \texttt{nr\_drop}     & Avg.\# events generated by the task but failed to collect \\  
& \texttt{latency}      & Avg. consumer-producer distance                            \\  
& \texttt{availability} & Size of available provenance buffer                   \smallskip\\ 
\bottomrule

\end{tabular}
}
\caption{Task and provenance context features collected by \sys.}
\label{tab:feature}
\end{table}

This provides a feasible baseline setting.
However, it still fails to provide a fine-grained budget control and thus, leads to sub-optimal performance. Building upon this baseline, we generalize this approach for a system configured with $N$ queues. With the intuition that exponential waiting times help queues being consumed in an ordered and predictable manner, let the resulting waiting time be denoted as $\hat{t}_\infty$, we set the waiting time for each queue exponentially as follows:
\begin{equation}
    \hat{t}_i=(\hat{t}_\infty)^{(i+0.5)/N},~1<i\leq N
\end{equation}

An example of a scheduling with exponential waiting time $\hat{T}=\{2,4,8\}$ is illustrated in ~\autoref{tab:queues}. Within 12 time slots, the queues receive \mao{\revlabel{typo1}16.7\%, 16.7\%, 25\%, and 41.7\%} CPU share, respectively.
The example shows that, with exponential waiting times, queues are consumed in a fine-grained and predictable manner.

\subsection{\sys Learning Component}\label{sec:adaptation}
As shown in \autoref{fig:overview}, a dispatch cycle consists of a
task execution, context update, and network-based rescheduling.
The task is then placed in a queue and waits to be run.
In this section, we discuss how \sys utilizes reinforcement learning to \fixme{empirically} achieve zero event loss and high throughput with the backbone based on dispatch cycles.

\subsubsection{Predicting task behaviors: RL in \sys}\label{sec:rl}
Due to the dynamic nature of provenance workloads, traditional heuristic-based schedulers are unable to adapt to the diverse behaviors of provenance tasks.
To address this, \sys employs reinforcement learning (RL) to predict task behaviors and make intelligent scheduling decisions.

\begin{table}
\centering
\resizebox{\columnwidth}{!}{
\begin{tabular}{ccccccccccccc}
\toprule%
Time & 1 & 2 & 3 & 4 & 5 & 6 & 7 & 8 & 9 & 10 & 11 & 12 \\
\cmidrule[0.6pt](lr{0.125em}){1-13}
$h_4$
& 1 & $\frac{9}{8}$ & \sqbox{$\mathbf{\frac{5}{4}}$} & - & - & - & - & - & - & - & 1 & \sqbox{$\mathbf{\frac{9}{8}}$} \\
\\[-2ex]
$h_3$& 2 & \sqbox{$\mathbf{\frac{9}{4}}$} & - & - & - & 1 & \sqbox{$\mathbf{\frac{5}{4}}$} & - & - & - & \sqbox{\textbf{1}} & -\\ 
\\[-2ex]
$h_2$&\sqbox{\textbf{4}} & - & 1& \sqbox{$\mathbf{\frac{3}{2}}$} & - & \sqbox{\textbf{1}} & - & \sqbox{\textbf{1}} & - & \sqbox{\textbf{1}} & - & 1 \\ 
\\[-2ex]
\cmidrule[0.6pt](lr{0.125em}){1-13}
Result   &  
2 & 3 & 4 & 2 & 1 & 2 & 3 & 2 & 1 & 2 & 3 & 4 \\  
\bottomrule
\end{tabular}
}
\caption{Example scheduling with $\hat{T}=\{2,4,8\}$. Assuming the CPU has been idle for 8 time slots at start, the table gives the hungry factors and the final scheduling decision on which queue to consume.}
\label{tab:queues}
\end{table}

The overall RL training process is to maximize the expected reward by learning the optimal policy, i.e., the optimal action to take in a given state.
In the context of \sys, the state $S$ is defined as the context of the task, which includes the task's behavior and the provenance system's state.
The action $A$ is the decision of which queue to put the task in, and the network is the neural network that predicts the action.
Specifically, for the $i$-th queue $q_i$, we define the action table \(A=\left\{a_i,\ldots, a_N\right\} \), where each action $a_i$ puts the task into $q_i$:
\begin{equation}
    a_i:(\text{task}\rightarrow \text{queue}_i)
\end{equation}
\mao{\revlabel{why-dqn}Since \sys has a small state and action space, a lightweight Deep Q-Network (DQN)\cite{dqn} is sufficient for predicting scheduling decisions.
}
Let the weight parameters of the DQN be $\mathit{\theta}$.
The DQN predicts Q-values, the reward, for each action and selects the $k$-th queue with the highest reward:
\begin{equation}
    k=\text{argmax}_i\{\mathit{\theta}(S)\}
\end{equation}
The reward $r$ is the sum of $r_c$ and $r_p$,
\(    r=r_c+r_p \), reflecting both the scheduler's completeness goal using \(r_c\) and performance goal with \(r_p\), as discussed in the following sections.

\smallskip
\noindent\textbf{Optimizing provenance availability~~}
During training for provenance completeness, \sys aims to zero out event loss in the simulated runs.
At cycle $t$, let the total number of events be $E$ and the number of dropped events be $E^d$, we define the provenance reward $r_c$ as follows:
\begin{equation}
    r_c=-{E^d_{t+1}}/{E_{t+1}}
\end{equation}
The network is trained to minimize the event loss in the next cycle, \(t+1\).
A trivial policy could drive event loss to zero by consistently moving producer tasks to low-priority queues. \revlabel{typo3}The throughput reward below discourages this overly conservative behavior.

\smallskip
\noindent\textbf{Optimizing throughput~~}
The performance goal of \sys is to maximize system throughput by optimizing CPU utilization.
\mao{\revlabel{idle-reward}We define idleness as the number of idle slices observed before the task runs.}
Let $C_t$ be the idleness in the current dispatch cycle and $C_{t+1}$ be the idleness in the next cycle of the same task.
The utilization reward $r_p$ is defined as:
\begin{equation}
    r_p={C_{t}}/{(C_{t}+C_{t+1}+1)}
\end{equation}
This reward encourages the scheduler to reduce idleness in the next dispatch cycle by lowering $C_{t+1}$.
\mao{\revlabel{dual-reward}The dual-reward design prioritizes event-loss reduction because $r_c\in(-\infty,0]$, while $r_p\in[0,1)$.}

\subsubsection{Reducing the overhead}\label{sec:overhead}
To reduce the cost of neural network inference, we introduce a $\delta$-function to reduce unnecessary scheduling decisions given that \sys's
network inference could be more computationally expensive than existing schedulers.
With this optimization, the scheduler first compares the current task context \(f_{t} \) with its previous one \(f_{t-1} \) and then, only performs the neural network inference if the context has changed significantly, measured by a threshold $\delta$:
\begin{equation}
\begin{split}
|1-(f_{t-1}/{f_{t}})| &<\delta, ~\forall f\in \{\texttt{nvcsw}, E\}
\end{split}
\end{equation}

When comparing the context similarity, we consider two key factors: \texttt{nvcsw} (see \autoref{tab:feature}) and the average number of events ($E$).
A dramatic change in these metrics indicates workload shifts that require new scheduling actions.

\section{Evaluation}
\label{sec:eval}

We empirically evaluate \sys by answering the following research questions (RQs), with \textbf{[G3]} achieved by design with a theoretical evaluation in \autoref{sec:scheduler}:

\medskip
\noindent-~~\textbf{RQ1:} How well does \sys prevent event loss? \hfill\textbf{[G1]}

\smallskip
\noindent-~~\textbf{RQ2:} What is the runtime performance of \sys? \hfill\textbf{[G4]}

\smallskip
\noindent-~~\textbf{RQ3:} Is runtime bounded in the worst case? \hfill\textbf{[G2]}

\smallskip
\noindent-~~\textbf{RQ4:} What is the scheduling cost of \sys? \hfill\textbf{[G5]}

\smallskip
\noindent-~~\textbf{RQ5:} How do different settings affect \sys?

\mao{
\smallskip
\noindent-~~\textbf{RQ6:} How is CPU time shared between super producers and normal tasks?\revlabel{rq6}
}

\begin{figure*}
\captionsetup[subfigure]{justification=centering}
    \centering
    \subfloat[With over 1M syscalls/s, \sys does not lose events in cases where others do.]{
    \includegraphics[height=5.2cm]{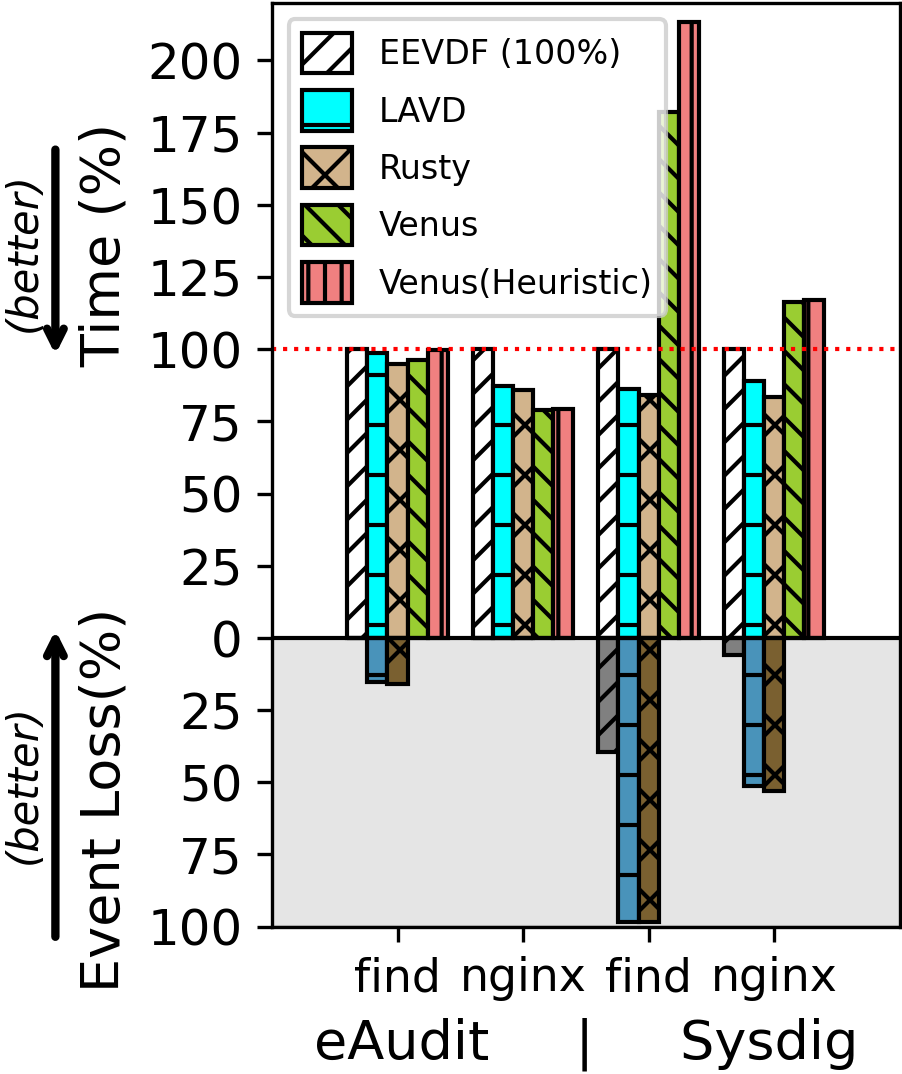}
    \label{fig:macrobench-heavy}
    }
    \subfloat[With less than 1M syscalls/s, \sys performs no worse (i.e., offers comparable  ~performance) compared to existing approaches.]{
    \includegraphics[height=5.2cm]{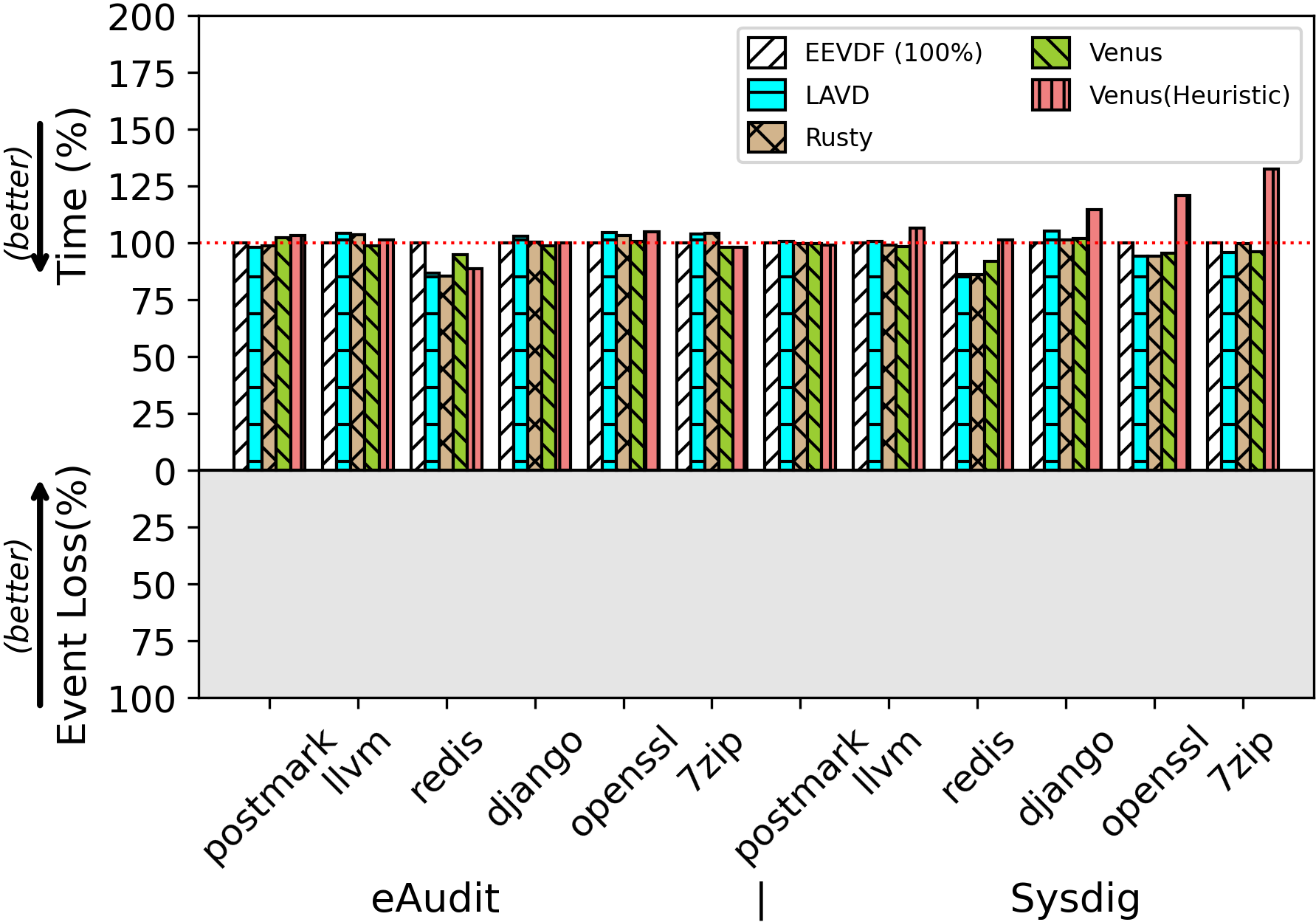}
    \label{fig:macrobench-normal}
    }
    \subfloat[\mao{Stress-test performance with Sysdig}]{
    \includegraphics[height=5.2cm]{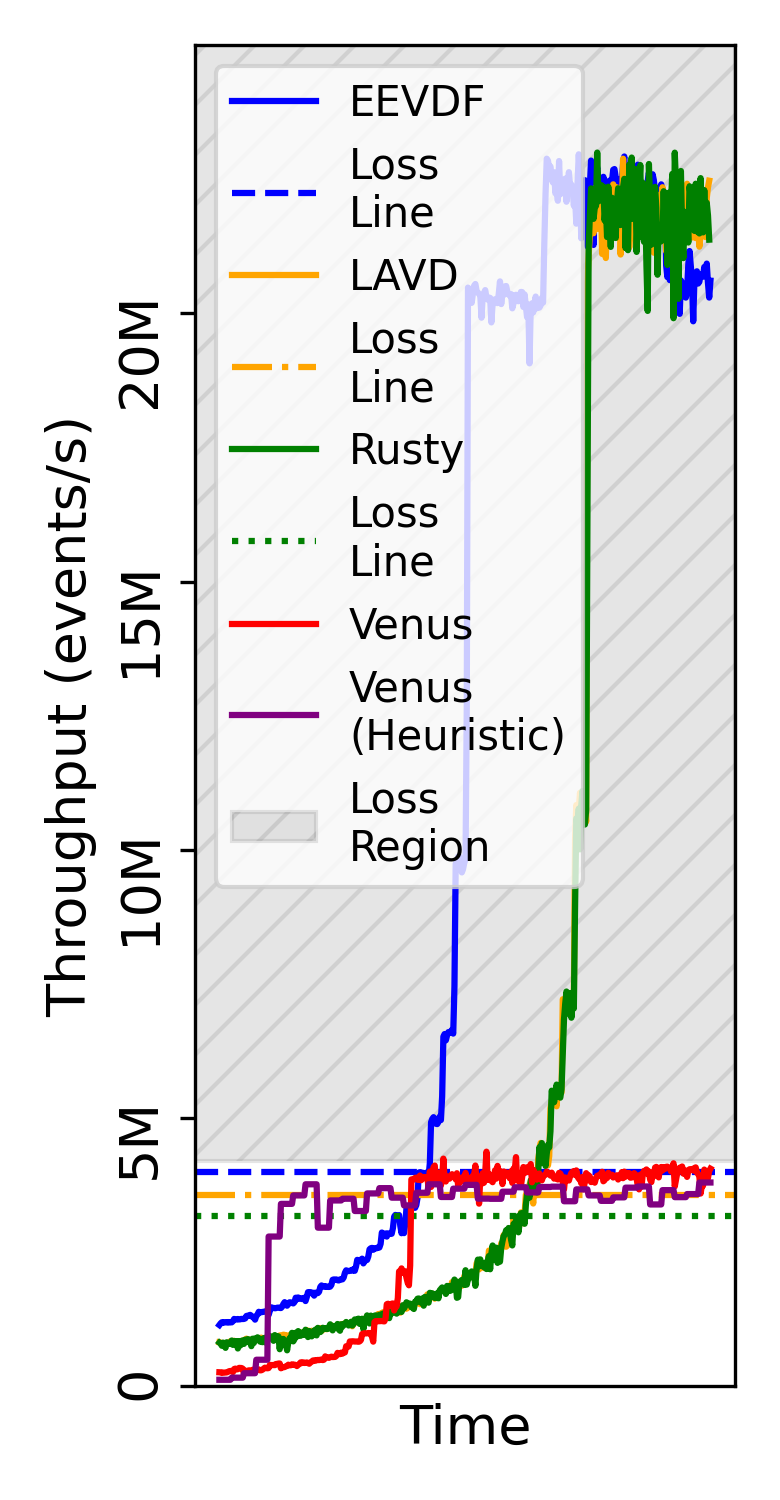}
    \label{fig:syscall-sysdig}
    }
    \subfloat[\mao{Stress-test performance with eAudit}]{
    \includegraphics[height=5.2cm]{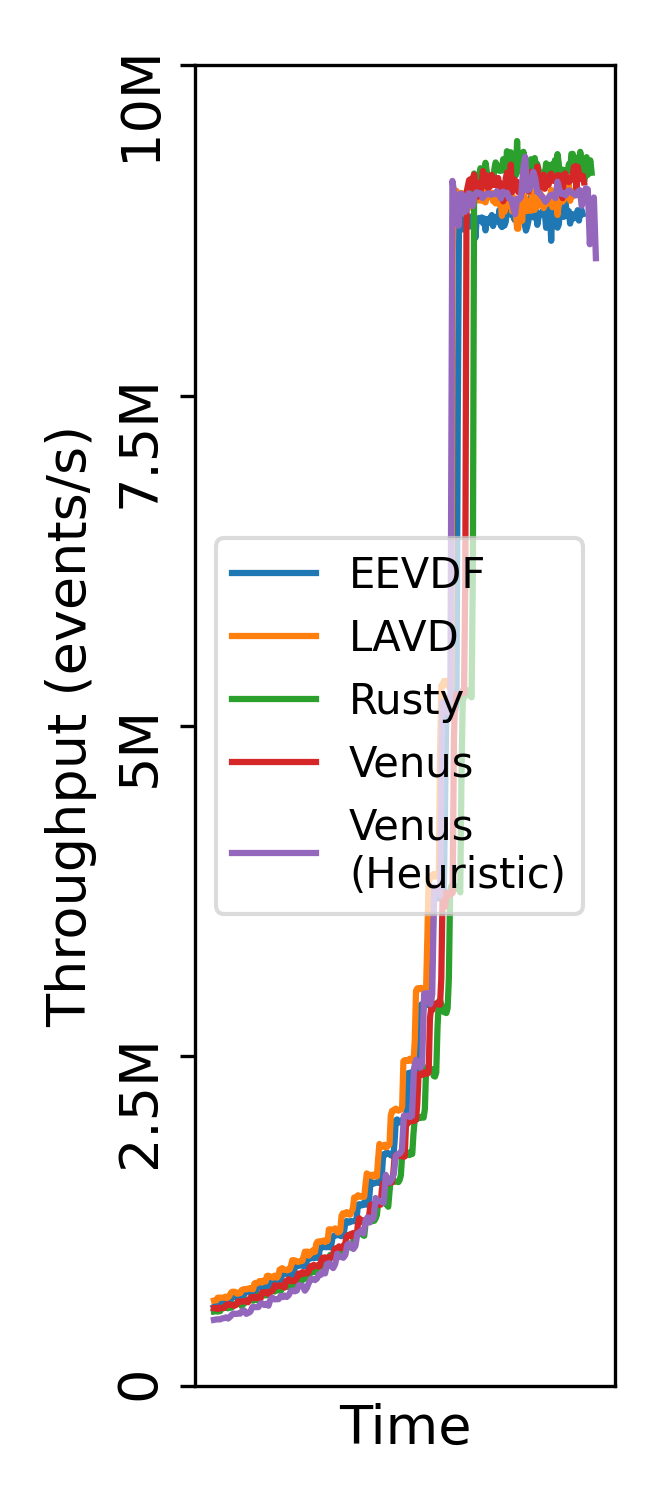}
    \label{fig:syscall-eaudit}
    }
    \caption{Performance of \sys on macrobenchmarks compared to EEVDF (baseline of 100\%)
    \mao{and under stress tests\revlabel{stress-figure}}}
    \label{fig:macrobench}
\end{figure*}

\subsection{Evaluation Setup}
\label{sec:setup}

Our prototype 
scheduler and its userspace daemon are written in C and compiled using LLVM 18.
Training and quantization use Python 3.10 and PyTorch 2.4.0. 
Communication between the scheduler and the user-space daemon is facilitated through sockets. We conduct experiments 
on an Ubuntu 24.04 QEMU virtual machine with an 8-core Intel 12900H processor, 32GB of DDR5 memory, and NVMe SSD. The system runs a Linux 6.10 kernel with \texttt{sched\_ext} support.

\smallskip
\noindent\textbf{Provenance systems~~}
To demonstrate the generality of \sys, we evaluate \sys using two open-source provenance systems, \textbf{Sysdig}\cite{sysdig} and \textbf{eAudit}\cite{eaudit}.
We utilize the same system call settings as eAudit, allowing the auditing of a total of 80 system calls.
We do not evaluate NoDrop\cite{nodrop} because its support is limited to legacy Linux kernels (v4.15) and is incompatible with modern 6.x kernels. 

\smallskip
\noindent\textbf{Schedulers~~}
We compare \sys with three state-of-the-art schedulers.
\sys operates on the \texttt{sched\_ext} framework, which requires Linux kernel version 6.8 or later. Thus, we use the \textbf{EEVDF}\cite{eevdf} scheduler as the baseline, which is Linux's default scheduler after kernel version 6.6. (We do not use CFS since one cannot revert to it with a recent kernel.) We also compare with \textbf{Rusty}\cite{rusty} and \textbf{LAVD}\cite{lavd}, the most popular schedulers from \texttt{sched\_ext} collections.

\smallskip
\noindent\textbf{Benchmarks~~}
To evaluate the performance and resilience of \sys, we select a diverse set of macrobenchmarks that utilize different kinds of system resources, performing CPU, I/O, and network activities. The benchmarks are summarized in ~\autoref{tab:benchmarks} along with their system call rates,
which we use as a proxy for provenance workload intensity.
Among all the tests, \textbf{7Zip} and \textbf{OpenSSL}
perform intensive compression algorithms and cryptographic operations. \textbf{Postmark} generates heavy file operations. \textbf{LLVM} test compiles the LLVM project and \textbf{find} test\cite{eaudit} spawns lots of processes to scan the \texttt{/usr/} directory simultaneously.
Representative server benchmarks like \textbf{Django}, \textbf{Redis} and \textbf{Nginx} are added to the above tests; they incur heavy network and IPC activities.
We use the same settings as NoDrop\cite{nodrop} in 7Zip, OpenSSL, Django, Redis and Nginx tests, and the same settings as eAudit\cite{eaudit} in Postmark and find tests.
Except for the purely computational 7Zip and OpenSSL tests, all tests incur substantial provenance events, stressing both \sys and provenance system.
\noindent Additionally, we introduce the \textbf{lmbench}\cite{lmbench} microbenchmark suite to measure low-level system performance.
We report the event loss ratio and the runtime of each benchmark under different schedulers and provenance systems. All reported results are averages of 10 attempts.

\begin{table}
\centering
\resizebox{\columnwidth}{!}{
\begin{tabular}{L{2.2cm}C{2.2cm}R{1.7cm}R{1.7cm}}
\toprule%
\multirow{2}{*}{\bfseries{Benchmarks}} & \multirow{2}{*}{\bfseries{Concurrency}} & \multicolumn{1}{c}{\multirow{2}{*}{\shortstack{\bfseries{Generic I/O} \\ (syscalls/s)}}} & \multicolumn{1}{c}{\multirow{2}{*}{\shortstack{\bfseries{Network} \\ (syscalls/s)}}} \\ \\
\cmidrule[0.6pt](r{0.125em}){1-4}%

\textbf{7Zip\cite{7zip}}               & Multi-core  & / & / \\ [-2ex]\\
\textbf{OpenSSL\cite{openssl}}            & Multi-core  & / & / \\ [-2ex]\\
\textbf{Postmark\cite{postmark}}           & Single-core & 316,657 & / \\ [-2ex]\\
\textbf{LLVM\cite{llvm}}               & Multi-core  & 14,892 & / \\ [-2ex]\\
\textbf{find\cite{eaudit}}  & Multi-core  & 1,098,652 & / \\ [-2ex]\\ 
\textbf{Django\cite{django}}             & Single-core & 4,257 & 71 \\ [-2ex]\\
\textbf{Redis\cite{redis}}              & Multi-core  & 298,124 & 122 \\ [-2ex]\\
\textbf{Nginx\cite{nginx}}              & Multi-core  & 915,771 & 304,856 \\ [-2ex]\\

\bottomrule
\end{tabular}
}
\caption{Benchmarks for evaluating \sys. Syscall speeds are average values under Sysdig and \sys. Generic I/O syscalls include \texttt{openat}, \texttt{read}, \texttt{write}, \texttt{writev} and \texttt{close}. Network-specific syscalls include \texttt{recvmsg}, \texttt{recvfrom}, \texttt{accept4}, \texttt{socket} and \texttt{connect}.}
\label{tab:benchmarks}
\end{table}

\smallskip
\noindent\textbf{Queues~~}
We configure four queues: one primary queue and three additional queues. The consumer, which receives and processes events, is pinned to the primary queue to facilitate the training process.
The determined $\hat{t}_\infty$ is $5e^5$ ns.

\vspace{4pt}
\noindent\textbf{Neural networks~~}
For each provenance system, we train an independent neural network, which is a two-layer fully connected model activated by ReLU \cite{relu}. The hidden layer consists of 256 neurons, with an input size of 11 and an output size of 4, resulting in a total size of 11 $\times$ 256 $\times$ 4. Key hyperparameters include a discount factor $\gamma=0.9$, a soft update coefficient $\tau=0.005$, and a learning rate of $1e^{-4}$.

\fixme{
To show the effectiveness of the NN, we also evaluate \sys(Heuristic), which keeps the same queues and waiting times, but uses a task's recent syscall rate as the indicator of its pressure on the provenance system. Let $r$ be the number of syscalls a task issued in the most recent window of length $\delta$, divided by $\delta$. Given thresholds $0=\theta_1<\theta_2<\theta_3<\theta_4$, the task is placed into queue $q_k$ with $k=\max\{i \mid r\ge\theta_i\}$, i.e., tasks with higher syscall rates are placed into queues with longer waiting times. For each provenance system, the thresholds are tuned on the test workloads by binary search, from $\theta_4$ down to $\theta_2$: since a lower threshold throttles more tasks and increases runtime, each $\theta_i$ is set to the largest value below $\theta_{i+1}$ that causes no event loss.
}

\vspace{4pt}
\noindent\textbf{Training~~}
During the training process, \sys continuously collects statistics of various workloads that include daily remote development, coding with VS Code and web browsing with Firefox, supplemented by periodic (e.g., every 3 mins) controlled stress using a lightweight variant of the find test with limited concurrency to prevent overfitting.
The model will not see test workloads in \autoref{tab:benchmarks}. Once the model converges, the weights are frozen for testing.

\mao{
\revlabel{training}
We train the model for 4000 steps with a batch size of 400, and stop once the loss has not improved for 2000 consecutive steps. Across 10 CPU-only runs, training takes 80.6 minutes on average, with a minimum of 42.8 minutes and a maximum of 130.5 minutes. This variation is expected due to the exploratory nature of reinforcement learning. 
}
\fixme{Across these runs, we observe that the loss stays above 50 and fluctuates during the first 50--80 steps, as workloads shift and new workloads enter the system (e.g., the first invocation of the lightweight find test). After this initial phase, the loss decreases smoothly to around 3 and remains stable for the rest of the training, even when new applications are introduced afterwards (e.g., opening Firefox at step 400 or compiling a project at step 670). This indicates that loss stabilizes quickly once the model has seen a sufficiently diverse set of task behaviors.}

\subsection{RQ1: Efficacy of \sys}
\label{sec:effective}

A critical aspect of \sys is its effectiveness in preventing provenance event loss. The grey area of ~\autoref{fig:macrobench} shows the event loss stats.
For all tests, \sys demonstrates zero observed event loss for both Sysdig and eAudit, showcasing its ability to handle high-frequency event streams reliably.

In contrast, in the find test, EEVDF incurred a significant event loss rate of 39.34\% with Sysdig, while maintaining no event loss with eAudit. LAVD and Rusty both exhibited severe event loss. LAVD reports event loss rates of 98.53\% for Sysdig and 15.11\% for eAudit, whereas eAudit is designed to achieve the completeness of events. Rusty shows nearly identical results. These results indicate that existing schedulers fail to prioritize  provenance consumers sufficiently, underscoring \sys's robust handling of diverse workloads by comparison.
For the Nginx test, EEVDF reports a loss rate of 5.9\% for Sysdig. LAVD and Rusty also fail to provide event completeness, reporting loss rate of 51.42\% and 52.93\% respectively, with Sysdig. However, \sys achieved zero event loss, reaffirming its ability to sustain complete provenance data in various conditions.
All schedulers successfully avoided event loss for both Sysdig and eAudit for other benchmarks, which is expected since they are relatively lightweight compared to Nginx and find test. Overall, \sys consistently achieves zero event loss across diverse benchmarks and provenance systems, outperforming state-of-the-art schedulers like EEVDF, LAVD, and Rusty.

\smallskip
\takeaway{Takeaways}{
\sys \fixme{observes} zero event loss across all test scenarios, even when other schedulers lose $>98\%$ events.
}

\subsection{RQ2: Performance of \sys}
\label{sec:performance}

We consider \sys's macrobenchmark, microbenchmark, \mao{\revlabel{stress-setting}and stress-test} performance.

\vspace{4pt}
\noindent\textbf{Macrobenchmarks~~}
As shown in ~\autoref{fig:macrobench}, \sys has the best performance in 10 of 16 tests.
We observe that find and Nginx tests are the most provenance-demanding tests. In the two tests, although Rusty and LAVD are faster than \sys in three cases, they \revlabel{typo4}drop provenance events.
With eAudit, \sys is faster than EEVDF, LAVD and Rusty in terms of average performance. With Sysdig, 
\sys has average overhead of 49.24\% compared to EEVDF, 70.96\% compared to LAVD and 78.06\% compared to Rusty. The overhead is high because the event loss ratio is high (e.g., 39.34\%, 98.53\% and 98.54\%, respectively).

For Django and LLVM tests, \sys is on average 0.42\%, 3.46\% and 1.53\% faster than EEVDF, LAVD and Rusty. This showcases efficient resource allocation of \sys when handling daily workloads. In Postmark, \sys is 0.86\%, 1.55\%, 1.76\% slower, respectively. Such overhead is mainly caused by the scheduling cost and does not pose a significant concern in real scenarios. For Redis, \sys is 6.63\% faster than EEVDF but 8.07\% and 8.90\% slower than LAVD and Rusty. This is because \sys is designed for general provenance scenarios rather than being a latency-critical scheduler like LAVD and Rusty.
For computational workload including 7Zip and OpenSSL, \sys dominates other schedulers, achieving 2.42\%, 2.62\% and 1.82\% faster than the other three schedulers. This further suggests that \sys improves task behavior recognition and resource utilization compared to state-of-the-art schedulers. In conclusion, all tests under \sys finish in a reasonable time without provenance event loss.

\fixme{Compared to \sys (Heuristic), \sys is 12.24\% faster on average with Sysdig and 0.79\% faster on average with eAudit, which indicates that the RL component of \sys is effective in improving runtime performance. Simple heuristic rules cannot capture the complex and dynamic nature of diverse workloads, leading to over-throttling in this scenario, while RL can adapt to changing conditions and optimize scheduling decisions based on observed patterns.}

\mao{
\revlabel{stress-test}
\vspace{4pt}
\noindent\textbf{Stress Test~~} We implement a dummy syscall loop that writes to \texttt{/dev/null} and reads from \texttt{/dev/zero}. We gradually increase the syscall rate by decreasing the number of spins between loop iterations, and report the final event rate and loss statistics in~\autoref{fig:syscall-sysdig} and~\autoref{fig:syscall-eaudit}. For Sysdig, EEVDF, LAVD, and Rusty begin losing events at around 4.0M, 3.5M, and 3.1M events per second, respectively, whereas \sys and its heuristic version stabilize at 3.96M / 3.55M events per second without losing events. For eAudit, EEVDF, LAVD, Rusty, \sys(Heuristic) and \sys stabilize at 8.86M, 9.01M, 9.23M, 8.93M and 9.15M events per second, respectively, with no observed event loss.
}

\smallskip
\takeaway{Takeaways}{
Overall, \sys outperforms EEVDF, LAVD, and Rusty by 3.42\%, 0.90\% and 0.25\%, respectively, in macro workloads while incurring no event loss, 
and provides near-optimal performance in stress test among four schedulers.
}

\begin{figure}
    \centering
    \includegraphics[width=\columnwidth]{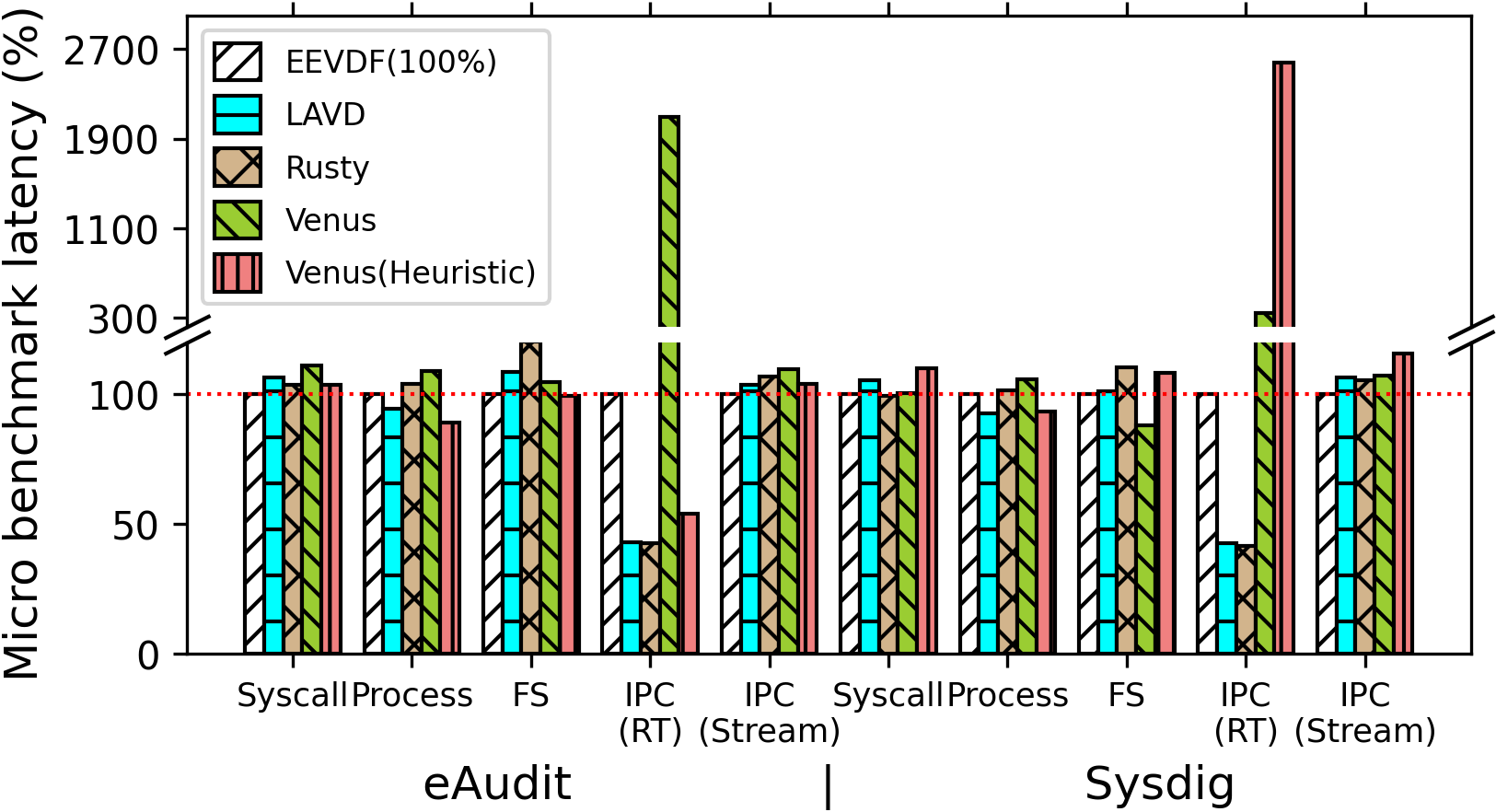}
    \caption{Average latency of microbenchmark tests compared to EEVDF (baseline of 100\%).}
    \label{fig:microbench}
\end{figure}

\noindent\textbf{Microbenchmarks~~}
We divide lmbench into four categories: system call, process, file system (FS) and inter-process communication (IPC). System call latency tests consist of NULL call, NULL I/O, stat, open/close file and signal install/handle. Process tests include fork, execve and shell tests. File system tests include file create and delete with size of 0k and 10k. IPC roundtrip tests include AF\_UNIX and pipe latency tests.
\fixme{To complement the IPC tests, we introduce a streaming variant that removes the reply dependency, allowing IPC to decouple sender and receiver.}

The results in~\autoref{fig:microbench} show that, compared to EEVDF and other schedulers, \sys does not incur noticeable overheads in system call, process, file system and streaming IPC tests. That is expected since many micro operations happen in one slice, and thus are not affected by the scheduler.
\sys shows higher overhead for roundtrip IPC; compared to the default EEVDF scheduler, \sys has 2.58× / 2.17× overhead with Sysdig on AF\_UNIX / pipe tests, respectively. For eAudit, the numbers are 20.12× / 19.79×.
The reason behind this is \sys's design using FIFO queues: when multiple tasks arrive at the same time, especially when related tasks (such as parent and child) are classified into different queues, the latency becomes high.

\fixme{\sys performs similarly to the heuristic-based baseline on Syscall, Process, FS tests and streaming IPC tests. On round trip IPC tests, \sys shows higher overhead with eAudit compared to heuristics, as eAudit suffers from less pressure than Sysdig and optimal heuristic threshold put all tasks to highest priority queue, exposing negligible overhead for roundtrips. Conversely, the heuristic introduces extra roundtrip IPC overhead for Sysdig as all workloads are classified to the least priority queue to avoid event loss.}

\smallskip
\takeaway{Takeaways}{
\sys performs similarly to other schedulers for most microbenchmark tests. While \sys introduces high latency in pipes and AF\_UNIX, the latency is still under 0.3ms and does not pose a significant concern to macro workloads, which \sys aims to improve.
}

\subsection{RQ3: Worst Case Analysis}
\label{sec:fair}

To evaluate the fairness and starvation resistance of \sys, we conduct a worst-case analysis for the scenario discussed in ~\autoref{sec:backbone}. We demonstrate \sys's robustness with six configurations with different queue and task settings, as shown in ~\autoref{tab:guarantee}.
Specifically, a pure computational task implemented as tight loops is introduced and used to saturate all higher-priority queues with 10 concurrent tasks.
In the lowest priority queue we adjust the number of tasks to observe the finish time.

\begin{table}
\centering
\resizebox{\columnwidth}{!}{
\begin{tabular}{cccccccccc}
\toprule%

\multirow{2}{*}{Setting}            & \multicolumn{4}{c}{\# of Tasks}  & \multicolumn{4}{c}{Waiting Time}         \\

\cmidrule[0.6pt](lr{0.125em}){2-5}%
\cmidrule[0.6pt](lr{0.125em}){6-9}

& Q1 & Q2 & Q3 & Q4 & Q1 & Q2 & Q3 & Q4 \\

\cmidrule[0.6pt](lr{0.125em}){1-9}

E2 & 10 & 2-100 & / & / & 1e2 & 5e5 & / & / \\
E3 & 10 &10  & 2-100 & / &1e2 & 7e3 & 5e5 & / \\
E4 & 10 &10 & 10 & 2-100 & 1e2 & 2e3& 3e4 & 5e5 \\ 
E5 & 10 &10 & 2-100 & / &1e2 & 2e3 & 4e4 & / \\
E6 & 10 &10 & 2-100 & / &1e2 & 3e4 & 6e6 & / \\
E7 & 10 &10 & 2-100 & / &1e2 & 8e4 & 7e7 & / \\

\cmidrule[0.6pt](lr{0.125em}){1-9}
RR & \multicolumn{8}{c}{5-104 Tasks} \\

\bottomrule
\end{tabular}
}
\caption{Setting of tasks and waiting times for the non primary queues for worst case analysis.}
\label{tab:guarantee}
\end{table}

Our results confirm that the lowest-priority task is consistently scheduled within the predicted bound. For \textbf{E2} to \textbf{E4}, the runtime ratios relative to the RR scheduler are 1.25× to 2.15×, 1.94× to 3.13× and 2.01× to 3.96×. The results indicate that the task in the lowest-priority queue
maintains predictable execution proportional to the number of queues configured,
validating the fairness guarantee enforced by \sys.
For \textbf{E5} to \textbf{E7}, the results are 2.24× to 2.94×, 2.75× to 4.01×, 6.69× to 16.75×. We observe that with longer waiting time configured, while the lowest priority queue gets a longer runtime, it is still bounded by \sys's queue-based mechanism (i.e., $10^4-10^6\times$).
Overall, the results indicate that in the worst case no starvation or indefinite postponement was observed.

\smallskip
\takeaway{Takeaways}{
\sys preserves queue-based fairness and bounded runtime for all tasks, even under intentionally adversarial settings.
}

\subsection{RQ4: Scheduling Cost of \sys}
\label{sec:cost}

The scheduling cost breaks down into \emph{inference cost} (percentage of time consumed by pre-processing and neural network inference for scheduling decisions) and \emph{maintenance cost} (collecting and updating features of each task and selecting the appropriate queue).
\autoref{tab:cost} shows the results.

\smallskip
\noindent\textbf{Inference cost~~} Across the benchmarks, \sys exhibits low inference costs. The most demanding benchmark, the Nginx test, has the highest inference cost of 2.44\%, and the find test has the highest inference frequency of 528 Hz. Multithreaded, I/O-heavy workloads in \sys often incur around 500Hz inference frequency per core such as find, LLVM, Nginx, and Redis. This is because these processes do not fully consume their slices and incur heavy scheduling decisions. The results also indicate computational workloads require $<$50 Hz of decision making. This is because they do not need to yield to other tasks.
At the same time, the results show that the average inference time is 50.16 $\mu s$. The lowest time, 37.33 $\mu s$ appears in lmbench, and the highest 60.31 $\mu s$ appears in 7Zip. This is because lmbench does not involve a lot of memory operations, and consequently the cache stays hot for the weights, reducing the overall weight accessing time. As a result, the inference cost is relatively lower.
\begin{table}
\centering
\resizebox{\columnwidth}{!}{
\begin{tabular}{crrr}
\toprule%
Metric       &Min &Max   & \multicolumn{1}{c}{\bfseries{Average}} \\
\cmidrule[0.6pt](lr{0.125em}){1-4}%

Inference Cost (\%)     &0.09 &2.75              &\textbf{1.18} \\
Inference Frequency Per Core (Hz)  &17.17 &528.00   &\textbf{294.77} \\ 
Inference Avg. Time ($\mu$s)   &37.33 &60.31           &\textbf{50.16} \\
Maintenance Cost (\%)   &0.017 &0.162             &\textbf{0.08}   \\
Delta Func Reduction (\%)    &0.04 &0.85       &\textbf{0.36}  \\ 
Delta Func Effect Ratio (\%)  &11.50 &98.46      &\textbf{46.92}  \\  
\cmidrule[0.6pt](lr{0.125em}){1-4}\\[-2ex]
\bfseries{Overall Cost (\%)}  &0.01 &2.44    &\textbf{0.82} \\  \\[-2ex]

\bottomrule
\end{tabular}
}
\caption{Scheduling cost of \sys.}
\label{tab:cost}
\end{table}

\medskip
\takeaway{Takeaways}{
The overall inference cost of \sys is low, under $0.5\%$ for typical scenarios and around $2\%$ for most demanding scenarios.
}

\smallskip
\noindent\textbf{Maintenance cost~~} The maintenance cost, representing the time spent updating task contexts and features, remains negligible across all benchmarks, ranging from 0.007\% in lmbench to 0.110\% in Nginx. With eBPF and \texttt{sched\_ext}, \sys acquires the features and contexts of current task directly from \texttt{task\_struct} and eBPF helper functions. Such design benefits 
\sys's task context management cost, even under high-frequency decision-making scenarios.

\smallskip
\noindent\textbf{Delta function~~}The delta function in \sys determines whether a task requires re-scheduling by evaluating changes in its context. We set $\delta=0.25$, which is equivalent to right shifting by two bits to simplify verifier reasoning and improve efficiency.
Results show that the delta function is particularly beneficial for Postmark, OpenSSL, 7Zip and lmbench, saving 49.95\% to 98.46\% scheduling decisions. This is because these long-running stable workloads show similar behavior across the whole time. In contrast, other workloads see a lower effective ratio ranging from 11.50\% to 37.96\% due to higher task context variability and shorter task lifecycle. For runtime, the delta function is capable of reducing the overall runtime by an average of 0.61\%, showing the effectiveness of reducing the overall computational cost of \sys.

\subsection{RQ5: Ablation Study}
\label{sec:ablation}

To evaluate the contributions of various design choices in \sys, we conduct an ablation study using Sysdig as the provenance system  and $\hat{t}_\infty=5e^5$ across two scenarios: feature configurations and queue configurations.

\vspace{4pt}
\noindent\textbf{Feature settings~~}
We investigate the effect of using different feature sets to train the neural network. 
We evaluate using task-only features (\textbf{F1}) and using provenance-only features (\textbf{F2}) from~\autoref{tab:feature} to train the neural network, and we compare the results with all features (\textbf{\sys}).

The results show that with \textbf{F1}, the system converges during training and is capable of preventing event loss under training workloads (i.e., 50 concurrent find workers). However, the scheduler is not robust enough and fails to work under unseen workloads (e.g., 288 workers). As a result,
it loses 55.46\% events.
\textbf{F2}
fails to adapt to even the training workloads, which leads to 23.63\%, 3.29\% and 29.74\% event loss in find, Postmark and Nginx tests. 
The results indicate that \sys's functionality relies on both kinds of features.

\vspace{4pt}
\noindent\textbf{Queue settings~~}
\mao{
\revlabel{queue-setting}
To evaluate the effect of queue configuration on system performance, we vary both the number of queues and the waiting time assigned to each queue.
We evaluate eight variants. \textbf{E1} to \textbf{E4} use one to four non-primary queues. \textbf{E5} to \textbf{E7} use three non-primary queues, as \sys does, but set the waiting times to smaller, larger, and significantly larger values than \sys, respectively, as shown in~\autoref{tab:guarantee}. We also compare a linear waiting-time setting (\textbf{E3L}) with \sys's exponential setting.}

\mao{
We observe that, when using only one non-primary queue or insufficient waiting time (\textbf{E1}/\textbf{E5}), the network converges during training but loses up to 60.29\%/69.34\% of events during testing, which is similar to that of \textbf{F1}.
When introducing two or four non-primary queues, or larger waiting times (\textbf{E2}/\textbf{E4}/\textbf{E6}), we find that the trained schedulers are capable of preventing event loss but lead to additional runtime.
Specifically, \textbf{E2}/\textbf{E4}/\textbf{E6} are 13.04\%/12.68\%/118.54\% slower in find test, 5.37\%/62.37\%/188.15\% slower in Nginx test and are overall 3.40\%/11.73\%/39.23\% slower than \textbf{\sys}. Thus, three non-primary queues is the best configuration since \textbf{E2} and \textbf{E6} give coarse-grained settings while the situation in \textbf{E4} is too complex for simple networks. Fine-grained control is more difficult to learn in either of these settings. For a very large waiting-time setting (\textbf{E7}), the scheduler does not converge within a 12-hour window because the extended waiting time excessively throttles the whole system. For linear waiting time \textbf{E3L}, we observe 
1.60\% extra runtime across all tests, indicating that exponential waiting time leads to easier learning patterns.
}

\begin{figure}
    \centering
    \includegraphics[width=\linewidth]{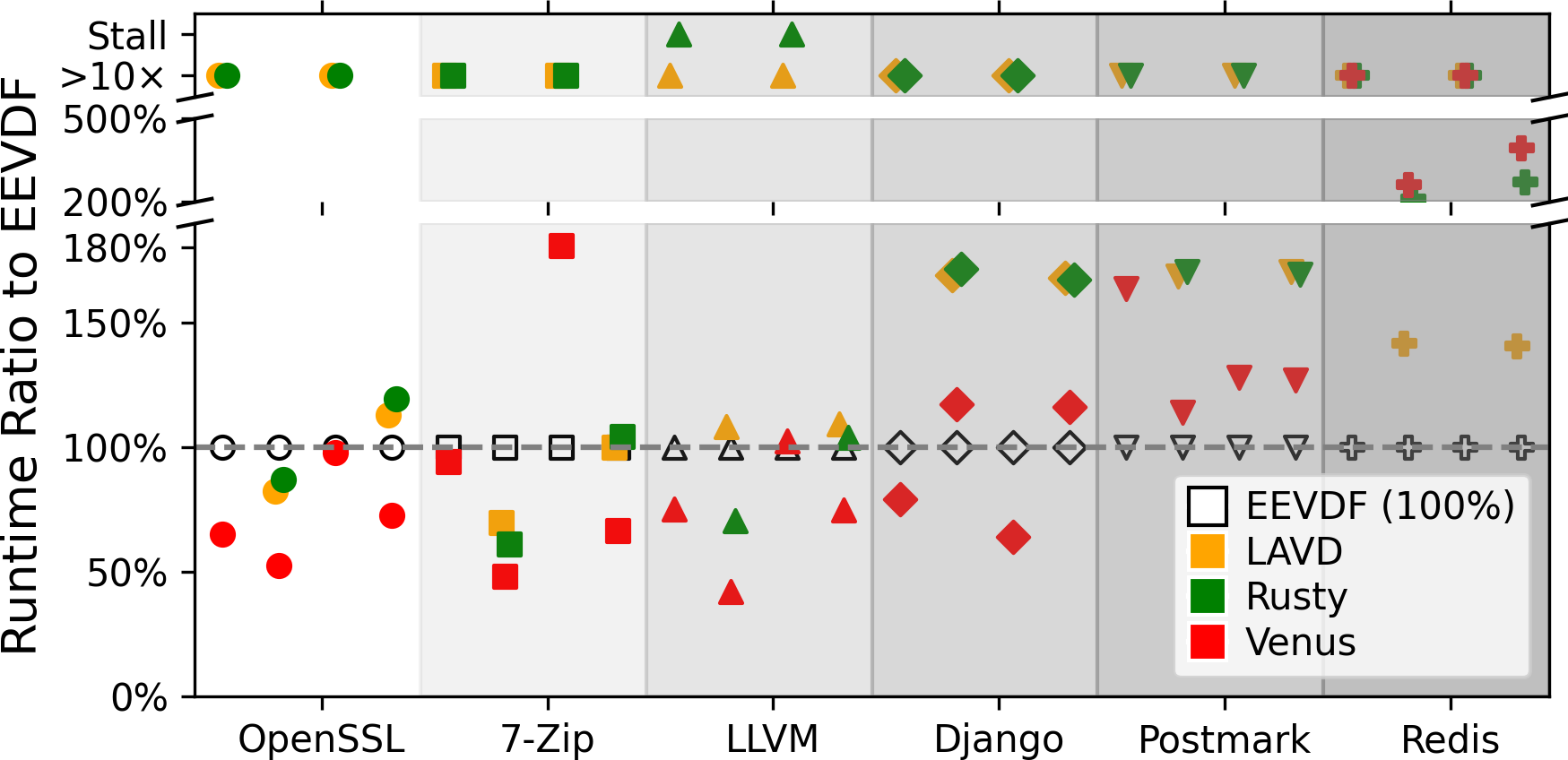}
    \caption{\mao{Performance of \sys on normal applications with background super producers running.
    Darker cells indicate heavier provenance workloads. Lower values indicate a larger CPU share for the foreground application.}}
    \label{fig:interference}
\end{figure}

\vspace{4pt}
\takeaway{Takeaways}{
Configuring \sys with exponential waiting times and four queues yields optimal performance.
}

\smallskip
\fixme{
\noindent\textbf{Deploying \sys In Practice~~}
These results suggest a three-step recipe for deploying \sys on a new host or with a new provenance system.
\textbf{First}, calibrate $\hat{t}_\infty$ using the two-queue stress procedure from \autoref{sec:waiting-time}, increasing the waiting time until the stress queue no longer induces event loss.
\textbf{Second}, grid-search the number of queues around four, using exponential waiting times. Too few queues are too coarse to separate super producers from normal tasks, while too many enlarge the action space beyond what a small network can learn reliably.
\textbf{Third}, freeze the network and deploy it to kernel space.
}

\mao{
\subsection{RQ6: Performance Interference}
\label{sec:interference}
\revlabel{interference}
To study CPU sharing between normal applications and super producers, we run \texttt{find} and Nginx as background super producers as both can saturate the provenance system. We then measure the runtime of the remaining six benchmarks. \autoref{fig:interference} reports results in the order find/Nginx with Sysdig, followed by find/Nginx with eAudit, for each benchmark.
}

\mao{Provenance-light workloads, including OpenSSL, 7Zip, LLVM compilation, and Django, often benefit from \sys because it suppresses background super producers. Compared with EEVDF, \sys gives these workloads up to 38.92\%, 2.63\%, 35.93\%, and 6.31\% more CPU share, respectively. In three cases, however, 7Zip and Django receive 44.76\%, 14.71\%, and 13.92\% less CPU share because \sys prioritizes overall provenance throughput and lets the super producers complete faster without event loss, as shown in~\autoref{fig:macrobench-heavy}.}

\mao{For provenance-heavy workloads, \sys may reduce the foreground application's CPU share under global overload prevention. Postmark, Redis with find, and Redis with Nginx receive 24.87\%, 55.93\%, and 98.11\% less CPU share, respectively, because these workloads also stress the provenance system. The exact ratio depends on Redis's multithreading behavior and the consumer throughput of each provenance system. Compared with LAVD and Rusty, \sys avoids severe slowdowns: both LAVD and Rusty incur more than $10\times$ slowdown for all tasks with the find background workload, and Rusty stalls the LLVM compilation test.}

\vspace{4pt}
\takeaway{Takeaways}{
Provenance-light tasks often receive more CPU share with \sys under background super producers.
}

\newcommand{\fsCandidates}{24}
\newcommand{\fsWindows}{2{,}303{,}372}
\newcommand{\fsWorkloads}{8}
\newcommand{\fsBaseAUC}{0.95}
\newcommand{\fsMaxAddAUC}{0.008}
\newcommand{\fsWithinCpuUtil}{0.00}
\newcommand{\fsRsqAvgProd}{0.998}
\newcommand{\fsdAUCAvgCons}{0.008}
\newcommand{\fsRsqNvcsw}{0.87}
\newcommand{\fsdAUCNvcswTs}{0.005}
\newcommand{\fsRhoAvgRuntime}{0.9994}
\newcommand{\fsdAUCVruntime}{-0.002}
\newcommand{\fsRsqDutyCycle}{0.83}
\newcommand{\fsRsqLatCri}{0.99}
\newcommand{\fsRsqEvtSz}{0.993}

\begin{table}[t]
\centering
\setlength{\tabcolsep}{4pt}
\caption{
    \fixme{Excluded candidate features.
\emph{Constant / Global}: the value does not differ across tasks (task var.: share of variance across tasks on the same CPU at the same time).
\emph{Duplicate}: the kept features in \autoref{tab:feature} predict the candidate ($R^2$), or it is strongly correlated with one ($|\rho|$).
\emph{No gain}: adding the candidate does not improve prediction of next-window event loss ($\Delta$AUC).}
}
\label{tab:discarded}
\resizebox{\columnwidth}{!}{%
\begin{tabular}{@{}lll@{}}
\toprule
\textbf{Feature} & \textbf{Description} & \textbf{Evidence} \\
\midrule
\texttt{ringbuf\_sz}        & Provenance buffer capacity       & Constant (fixed size) \\
\texttt{cpu\_util}          & Utilization of the task's CPU    & Global (task var.\ \fsWithinCpuUtil) \\
\texttt{avg\_produce}       & Task's event production rate     & Duplicate ($R^2 = \fsRsqAvgProd$) \\
\texttt{evt\_sz}            & Avg.\ size of the task's events  & Duplicate ($R^2 = \fsRsqEvtSz$) \\
\texttt{lat\_cri}           & Latency criticality (LAVD\cite{lavd})       & Duplicate ($R^2 = \fsRsqLatCri$) \\
\texttt{avg\_runtime}       & Avg.\ CPU time per run           & Duplicate ($|\rho| = \fsRhoAvgRuntime$, \texttt{nvcsw}) \\
\texttt{avg\_freq\_runtime} & Frequency-scaled avg.\ runtime   & Duplicate (fixed CPU freq) \\
\texttt{total\_nvcsw}       & Total voluntary context switches & Duplicate ($R^2 = \fsRsqNvcsw$) \\
\texttt{duty\_cycle}        & Fraction of time the task runs   & Duplicate ($R^2 = \fsRsqDutyCycle$) \\
\texttt{avg\_consume}       & Consumer's event processing rate & No gain ($\Delta$AUC $\fsdAUCAvgCons$) \\
\texttt{nvcsw\_ts}          & Time of last voluntary switch    & No gain ($\Delta$AUC $\fsdAUCNvcswTs$) \\
\texttt{vruntime}           & Virtual runtime (CFS/EEVDF)      & No gain ($\Delta$AUC $\fsdAUCVruntime$) \\
\texttt{avg\_event\_xxx}    & Per-category event counts        & Too costly to train \\
\bottomrule
\end{tabular}
}
\end{table}

\section{Related Work and Discussion}
\label{sec:related}

\noindent\textbf{Schedulers~~} In modern Linux kernels, CFS\cite{cfs} and EEVDF\cite{eevdf} replace \texttt{O(1)} scheduler\cite{o1scheduler} with finer granularity and better fairness. Similar schedulers~\cite{karma, rusty} provide generic efficiency and fairness among all tasks.
Domain-specific efforts explore gaming~\cite{lavd}, energy~\cite{nest, liang2023energy} and micro operations~\cite{mcclure2022efficient, lin2024fast}. Recently, scheduler frameworks~\cite{ghost,plugsched, schedext, syrup, skyloft} enable user-defined scheduling; \sys adopts~\cite{schedext}.

\vspace{2pt}
\noindent\textbf{Provenance~~} Early provenance systems rely on instrumenting the OS kernel\cite{camflow,hifi,lpm,kcal}. Recent systems are mainly built with eBPF\cite{sysdig,tracee,tetragon,provbpf,eaudit}.
\mao{\revlabel{citation}Researchers also explore real-time auditing\cite{rtss25, audit-real-time} and hardware-based solutions\cite{dpuaudit,hardlog}.}
Some systems\cite{mpi,protracer,alchemist} also implement high-level semantic provenance. However, provenance completeness is less studied, and existing works on buffer design\cite{eaudit} and resource isolation\cite{nodrop} do not ensure all desired properties as discussed in~\autoref{sec:moti}.
While \sys addresses these critical challenges, broader provenance integrity remains complementary, such as incorporating orthogonal threat detection or
tamper-proofing techniques\cite{paccagnella2020logging,hoang2022faster,sealfs,sealfsv2,omnilog,hitchhiker}.

\vspace{2pt}
\noindent\textbf{Kernel-space machine learning~~} Kernel-space ML is bringing adaptive algorithms to scheduling\cite{mlbb}, storage\cite{linnos,kmlib}, networking\cite{ebpf-network,ebpf-network2,liteflow,kmod-bbr}, IDS\cite{ebpf-ids,ebpf-ids2} and compartmen-talization\cite{ebpf-compartmentalization}. Existing works use floating point\cite{kmlib}, quanti-zation\cite{liteflow} and API remoting\cite{lake} for computation. The closest work to \sys is MLLB\cite{mlbb}, which improves a
function rather than implementing a complete scheduler.
Future improvements to \sys may come from larger models, better training strategies, or hardware acceleration\cite{lake}.

\mao{
\revlabel{feature-discussion}
\vspace{2pt}
\noindent\textbf{Features~~}
}
\fixme{
We considered \fsCandidates{} candidate features and kept the 11 in \autoref{tab:feature}.
Most excluded candidates (\autoref{tab:discarded}) restate a kept feature in another form, such as event rate or average runtime; the rest do not differ across tasks or describe the consumer rather than the scheduled task.
To confirm this, an instrumented build of \sys logged all candidates for \fsWindows{} decision windows across \fsWorkloads{} workloads under Sysdig, with tasks placed in random queues.
No excluded candidate improves next-window loss prediction by more than \fsMaxAddAUC{} AUROC over the kept set's \fsBaseAUC{}, which is within cross-validation variance.
Other provenance backends, hardware performance counters, or per-category event counts may offer useful signals, and we leave revisiting feature selection for new deployments to future work.
}

\fixme{
\vspace{2pt}
\noindent\textbf{Limitations~~}
\sys has two limitations.
First, \sys does not provide a formal guarantee of event completeness; its completeness properties are established empirically on the workloads, provenance systems, and platform we evaluate. Without such a guarantee, it would be preferable to preserve security-critical events (e.g., \texttt{execve} or \texttt{connect}). However, its completeness reward $r_c$ is defined over the aggregate drop ratio $E^d/E$ and therefore treats every event as equally important. A weighted reward that encodes per-event-type or per-subject importance is a natural extension, and we leave it to future work.
Second, our evaluation establishes generalization only within the scope of this paper. We have not, however, examined broader or domain-specific deployments, such as HPC clusters, database servers, or embedded hosts, whose task and event patterns may differ from those studied here. We open-source \sys together with its training pipeline and welcome validation, and retraining if needed, on such workloads.
}

\section{Conclusion}
We introduced \sys, a kernel-space learned provenance scheduler that
\mao{
\revlabel{conclusion-empirical}empirically prevents observed provenance event loss
}
under super producer threats that would otherwise hide malicious system events.
We showed existing solutions do not adequately defend against the threat because they either suffer from buffer overflows or fail to accommodate important functionalities.
We leveraged RL within the kernel's scheduler to learn the behavior of tasks and to optimize resource allocation to ensure completeness while being performant against other tasks.
We evaluated \sys's efficacy using Sysdig and eAudit to study the performance and cost and found that \fixme{\sys exhibited no event loss in all evaluated settings, reducing the risk that a super producer can hide malicious events from the provenance system.}

\section*{Acknowledgments}
\fixme{
We thank the anonymous reviewers for their insightful comments and constructive feedback, which substantially improved this paper.
This work is supported in part by the National Science Foundation under grants 2342250 and 2319944, by an Amazon Research Award (2024), and by an NVIDIA Academic Grant Program award (2025).
}




\bibliographystyle{plain}
\bibliography{refs}

%


\end{document}